\documentclass[iop,apj,tighten]{emulateapj}

\usepackage{graphicx}
\usepackage{url}

\shorttitle{The Recurrent Nova LMC 2009a}
\shortauthors{Bode et al.}

\begin{document}

\title{Pan-Chromatic observations of the Recurrent Nova LMC 2009\MakeLowercase{a} (LMC 1971\MakeLowercase{b})}

\author{M.~F. Bode$^{1}$, M.~J. Darnley$^{1}$, A.~P. Beardmore$^{2}$, J.~P. Osborne$^{2}$, K.~L. Page$^{2}$, F.~M. Walter$^{3}$,
J. Krautter$^{4}$, \\ A. Melandri$^{5}$, J.-U. Ness$^{6}$,  T.~J. O'Brien$^{7}$, M. Orio$^{8,9}$, G.~J. Schwarz$^{10}$, M.~M. Shara$^{11}$ and S. Starrfield$^{12}$}

\affil{$^{1}$Astrophysics Research Institute, Liverpool John Moores University, IC2, Brownlow Hill, Liverpool, L3 5RF, UK\\
$^{2}$Department of Physics and Astronomy, University of Leicester, Leicester, LE1 7RH, UK\\
$^{3}$Department of Physics and Astronomy, Stony Brook University, Stony Brook, NY 11794-3800, USA\\
$^{4}$Landessternwarte-Zentrum f\"{u}r Astronomie der Universit\"{a}t, K\"{o}nigstuhl, D-69117 Heidelberg, Germany\\
$^{5}$INAF-Osservatorio Astronomico Brera, via E. Bianchi 46, 23807, Merate, LC, Italy\\
$^{6}$XMM-Newton Science Operations Center, European Space Astronomy Center, E-28691 Villanueva de la Ca\~{n}ada, Madrid, Spain\\
$^{7}$Jodrell Bank Centre for Astrophysics, Alan Turing Building, University of Manchester, Manchester, M13 9PL, UK\\
$^{8}$Department of Astronomy, University of Wisconsin, 475 N. Charter Str., Madison, WI 53704, USA\\
$^{9}$INAF-Osservatorio di Padova, vicolo dellÕ Osservatorio 5, I-35122 Padova, Italy\\
$^{10}$American Astronomical Society, 2000 Florida Ave., Nw, Suite 300, DC 20009-1231, USA\\
$^{11}$American Museum of Natural History, 79th Street and Central Park West, New York, NY 10024, USA\\
$^{12}$School of Earth and Space Exploration, Arizona State University, Tempe, AZ 85287, USA}

\submitted{{\scriptsize Received 2015 September 28; accepted 2015 December 28}}
\journalinfo{The Astrophysical Journal, Draft version \today}

\begin{abstract}
Nova LMC 2009a is confirmed as a Recurrent Nova (RN) from positional coincidence with nova LMC 1971b. The observational data set is one of the most comprehensive for any Galactic or extragalactic RN: optical and near-IR photometry from outburst until over 6 years later; optical spectra for the first 6 months, and Swift satellite Ultraviolet and X-ray observations from 9 days to almost 1 year post-outburst. We find $M_V = -8.4\pm0.8_{\mathrm{r}}\pm0.7_{\mathrm{s}}$ and expansion velocities between 1000 and 4000 km s$^{-1}$. Coronal line emission before day 9 indicates shocks in the ejecta. Strengthening of He II $\lambda$4686 preceded the emergence of the Super-Soft Source (SSS) in X-rays at $\sim63-70$ days, which was initially very variable. Periodic modulations, $P=1.2$ days, most probably orbital in nature, were evident in the UV and optical from day 43. Subsequently, the SSS shows an oscillation with the same period but with a delay of 0.28P. The progenitor system has been identified; the secondary is most likely a sub-giant feeding a luminous accretion disk. Properties of the SSS infer a white dwarf (WD) mass $1.1 \mathrm{M}_\odot \lesssim M_{\rm WD} \lesssim 1.3 \mathrm{M}_\odot$. If the accretion occurs at constant rate, $\dot{\it{M}}_{\rm acc} \simeq 3.6^{+4.7}_{-2.5} \times 10^{-7}\,\mathrm{M}_\odot$\,yr$^{-1}$ is needed, consistent with nova models for an inter-eruption interval of 38 years, low outburst amplitude, progenitor position in the color-magnitude diagram, and spectral energy distribution at quiescence. We note striking similarities between LMC 2009a and the Galactic nova KT Eri, suggesting that KT Eri is a candidate RN. 
\end {abstract}

\keywords{galaxies: individual (LMC) --- stars: individual (Nova LMC 2009a, Nova LMC 1971b), novae, cataclysmic variables  --- white dwarfs}

\section{Introduction}

Classical Novae (CNe) are cataclysmic variable stars whose eruptions are due to a Thermonuclear Runaway (TNR) on the surface of a white dwarf in an interacting binary system \citep[see e.g.][]{sta08, bod10}. Recurrent Novae (RNe) are related to CNe, but have been seen to undergo more than one recorded eruption \citep[recurrence times $\sim 1-100$~years; see][]{dar14,dar15}, and may contain evolved secondary (mass-donating) stars \citep[see][for recent reviews]{anu08,schae10,dar12}. Recurrent Novae have been proposed as one of the candidates for the progenitors of Type Ia Supernovae 
(SNe; see e.g. \citealt{mao14} for a review of observational studies and \citealt{sta12,new14} for theoretical work on the potential for novae to give rise to SNIa explosions).

At present we know of a total of only ten RNe in the Galaxy. These appear to fall into three main groups, {\em viz.}: 

\begin{itemize}

\item{{\em RS~Ophiuchi/T~Coronae~Borealis} with red giant secondaries, consequent long orbital periods ($\sim$ several hundred days), rapid declines from eruption ($\sim0.3$ mag day$^{-1}$), high initial ejection velocities ($\ga 4000$ km~s$^{-1}$) and strong evidence of the interaction of the ejecta with the pre-existing circumstellar wind of the red giant \citep[from observations of optical coronal lines, non-thermal radio emission and hard X-ray development of RS~Oph; see papers in][]{eb08}.  The \citet{dar12} classification includes these systems in the red giant nova (RG-nova) group; }

\item{The more heterogeneous {\em U~Scorpii} group with members' central systems containing an evolved main sequence or sub-giant secondary with an orbital period much more similar to that seen in CNe (of order hours to a day), rapid optical declines (U~Sco itself being one of the fastest declining novae of any type), extremely high ejection velocities \citep[$v_{\mathrm{ej}} \sim 10,000$\,km\,s$^{-1}$, from FWZI of emission lines for U~Sco;][]{anu00} but no evidence of the type of shock interactions seen in RS~Oph post-eruption \citep[their post-eruption optical spectra resemble the `He/N' class of CNe;][]{wil92}.  The U~Sco group are members of the sub-giant nova (SG-nova) group;}

\item{{\em T~Pyxidis, IM~Normae} are again short orbital period systems and although their optical spectral evolution post-eruption is similar to one another, with their early time spectra resembling the `Fe~{\sc ii}' CNe, they show a very heterogenous set of moderately fast to slow declines in their optical light curves. The latter group of systems also seems to show ejected masses similar to those at the lower end of the ejected mass range for CNe with $M_{\rm ej} \sim 10^{-5}$ M$_{\odot}$ (i.e.\ one to two orders of magnitude greater than $M_{\rm ej} $ in the other two sub-groups of RNe noted above).  These systems, along with most CNe, populate the main-sequence nova (MS-nova) group.}

\end{itemize}

The short recurrence periods of RNe require high mass WD accretors and relatively high accretion rates \citep[e.g.][]{sta88, hen15, dar15}. Indeed, both RS~Oph and U~Sco appear to have WDs near the Chandrasekhar mass limit. The WD mass in both these systems has been proposed to be growing such that they are potential SN Ia progenitors \citep[see e.g.,][respectively]{sok06,sta88}, provided that the WD is originally of the CO rather than ONe sub-type.

The study of RNe is thus important for several broader fields of investigation including mass loss from red giants, the evolution of supernova remnants and the progenitors of Type Ia SNe. Progress in determining the latter association in particular, as well as exploring the evolutionary history of these close binary systems, is hampered by the relative rarity of Galactic RNe (although it is thought that many more RNe are lurking among the CNe population; see e.g.\ \citealt{pas14}, \citealt{sha15}, \citealt{wil14}). However, since the time of Edwin Hubble \citep[see e.g.][]{hub29}, CNe have been observed in extragalactic systems, in particular M31, with a total of over 40 nova candidates discovered to-date in the LMC\footnote{\url{http://www.mpe.mpg.de/\textasciitilde m31novae/opt/lmc/LMC\_table.html}} (\citealt{sha13}; see also \citealt{sha08} for a general review). Of those in the LMC, at least 2,  YY~Dor  \citep[1937, 2004;][]{mas14} 
and Nova LMC~1990b \citep[][previous eruption 1968]{sek90,sho91}  have been classified as recurrent \citep[with Nova LMC 2012 a suspected RN;][]{sch15} to which we now add Nova LMC 1971b/2009a, the subject of this paper.

\section{Observations of LMC 2009a}

Nova LMC 2009a was discovered on 2009 February 05.067 UT (which we take as $t = 0$) by \citet{lil09} at mag = 10.6 (unfiltered)  and located at RA $=5^{\mathrm{h}}40^{\mathrm{m}}44^{\mathrm{s}}\!.20$, Dec $= -66^{\circ}40^{\prime}11^{\prime\prime}\!\!.6$ ($\pm 0^{\prime\prime}\!\!.1$ in each coordinate, J2000). \cite{lil09} state that there was no object brighter than magnitude 14.0 at that position on 2009 January 31.065, but unfortunately we can find no other observations of the region to help tie down the outburst more precisely. For example, the region was not observed by the OGLE project around this time \citep{mro15}. It was immediately realised that this position lay close to that of the second nova to be discovered in the LMC in 1971. However, there was some confusion in the literature about the exact location of Nova LMC 1971b.

The discovery announcement of LMC 1971b \citep{gra71} reported a position of RA $= 5^{\mathrm{h}}40^{\mathrm{m}}\!.6$, Dec $= -66^{\circ}41^{\prime}$ (equinox 1975). The significant figures quoted suggest uncertainties of $11''$ in RA and $30''$ in Dec. \cite{cap90} quoted the same coordinates, but as epoch 1950. The coordinates in the General Catalogue of Variable Stars\footnote{http://www.sai.msu.su/gcvs/gcvs/} are RA $= 5^{\mathrm{h}} 40^{\mathrm{m}} 36^{\mathrm{s}}$, Dec $-66^{\circ} 42^{\prime}$ (equinox 1950). This agrees with Graham's coordinates to within the roundoff errors. \cite{sub02} later quoted the GCVS coordinates precessed to epoch 2000 (RA $=5^{\mathrm{h}}40^{\mathrm{m}}35^{\mathrm{s}}\!.22$, Dec $= -66^{\circ} 40^{\prime} 35^{\prime\prime}\!\!.2$). These are the coordinates in the SIMBAD catalog. We have measured the position of the nova on the discovery plate using the two-axis measuring engine of the Kitt Peak National Observatory and find it to be  RA $=5^{\mathrm{h}} 40^{\mathrm{m}} 44^{\mathrm{s}}\!.2$, Dec $= -66^{\circ} 40^{\prime} 11^{\prime\prime}\!\!.5$ (J2000). This is within $0^{\prime\prime}\!\!.1$  of our measured position of nova LMC 20009a, confirming this as a recurrent nova with inter-eruption period of 38 years or less (if other eruptions between 1971 and 2009 have been missed). It should also be noted that the location of the nova indicated in the published finding chart  \citep{sha00} again shows the wrong position.

From \cite{gra71}, the maximum, which was missed, occurred between between 1971 July 19 and 1971 August 16.4. The unwidened objective-prism spectroscopy consisted mainly of very broad Balmer emission lines and the almost equally bright band near 4600 Angstroms. No  photometry or light curve of the nova that we are aware of has been published and therefore more detailed comparison with Nova LMC 2009a is not possible.

\subsection{Optical and Infrared Photometry}

We obtained optical and near-IR (NIR) photometry with  the SMARTS 1.3m telescope at the Cerro Tololo Inter-American Observatory (CTIO), using the ANDICAM dual-channel imager from 2009 February 6 through  2015 April 2 (days 1 to 2247 after discovery).

ANDICAM employs a dichroic filter to enable simultaneous optical and NIR imaging. The optical detector is a Fairchild 447 $2048\times2048$ CCD. The $6^{\prime}\!.2\times6^{\prime}\!.2$  image is oversampled; data are taken with $2\times2$ pixel binning, for a $0^{\prime\prime}\!\!.369$~pixel$^{-1}$ plate scale. Bias subtraction and flat fielding are done by the SMARTS pipeline prior to data distribution. 

The NIR detector is a Rockwell $1024\times1024$ HgCdTe Array. The field is $2^{\prime}\!.4\times2^{\prime}\!.4$; the images are rebinned to $512\times512$ on the ground ($0^{\prime\prime}\!\!.274$~pixels). No other processing is done prior to data distribution\footnote{Full details are available at\\ \url{http://www.astronomy.ohio-state.edu/ANDICAM/detectors.html}}. We obtain 3 images dithered with a $20^{\prime\prime}$ throw. 

The optical and NIR images are obtained simultaneously, in pairs. On most nights we obtained full $BVR_{\mathrm{C}}I_{\mathrm{C}}, JHK$ coverage. However, due to problems with the filter wheel we were only able to obtain a single pair of exposures each night between 2009 February 6 and 16. Consequently, we have poor temporal coverage in the first 12 days, when the nova is decaying fastest. We obtained 2 observations per night on many nights between days 40 and 60 in order to search for optical modulation on the 1.2 day period seen in the {\it Swift} observations (see below). These observations were separated by $2-4$ hours, in order to mitigate against aliasing with a 1 day period.

We have obtained between 160 and 173 exposures in each band. Exposure times were increased from 2 to 300 seconds in the optical, and from 12 to 195 seconds in the NIR channel, as the nova faded. 

\subsubsection{Data reduction}

The optical images are delivered fully processed, but the NIR are not. For the latter, we subtracted the scaled dark image from each exposure, and divided by the normalized dome flat (dome flats are generally taken every third day; we selected the closest one). We median filtered  the three images to generate the local sky image, which we then subtracted from each image. Finally, we shifted and coadded the 3 images.

For both sets of data we used aperture photometry to measure instrumental magnitudes of the target and comparison stars in the field. 

We used a 7 pixel ($2^{\prime\prime}\!\!.6$ in the optical; $1^{\prime\prime}\!\!.9$ in the NIR) radius aperture for the photometric extractions. We selected 12 stars in the field to serve as optical comparisons. One of these  later proved to be unsuitable. None of the other stars appears to be variable at the 0.02 mag level. In the smaller NIR image we used 8 comparison stars. The background is the median level within the annulus from 17 to 27 pixels from the target.

To convert to apparent magnitudes, we determined the magnitudes of the comparison stars by using the Landolt standard fields observed on nights that the nova was also observed to establish the photometric zero-point. We assumed the standard extinction law and no color correction.  The standard deviations of the computed comparison magnitudes (which includes sky transparency as well as stellar variability) are generally $<0.04$~mag. The apparent magnitudes of the comparisons range between 14.4 and 18.7 at $B$, and from 13.8 to 16.2 at $I_{\mathrm{C}}$.

For the NIR calibration, we used the Two Micron All Sky Survey \citep[2MASS;][]{skr06} magnitudes of the 8 comparison stars to set the zero-point correction. We applied the standard CTIO extinction solution, but no color corrections.

\subsection{SMARTS Spectroscopy}

We obtained low dispersion spectra with the RC spectrograph on the SMARTS 1.5m telescope at CTIO. All observations were made by SMARTS service observers. The spectroscopic record extends from 2009 February 7
though 2009 December 26 (days 2 to 323, with the vast majority of the data before day 160), by which time it
had become too faint for the 1.5m.  We secured a total
of 56 spectra on 56 nights.
Observing conditions ranged from photometric to thick overcast.

The RC spectrograph is a long-slit instrument; the slit length subtends 5~arcmin on the sky. We used a $1^{\prime\prime}$ slit width. The detector is a Loral CCD. We used five spectroscopic set-ups, as detailed in Table~\ref{tbl-spsetups}.

\begin{deluxetable}{lll}
\tablewidth{\columnwidth}
\tablecaption{Spectrograph set-ups\label{tbl-spsetups}}
\tablewidth{0pt}
\tablehead{
\colhead{Set-up} & \colhead{Wavelength Range (\AA)} & \colhead{Resolution (\AA)}
}
\startdata
13/I    & $3146-9374$ & 17.2 \\
26/Ia  & $3660-5440$ & 4.3 \\
47/Ib  & $5650-6970$ & 3.1 \\
47/II   & $3880-4550$ & 1.6 \\ 
47/IIb & $4050-4720$ & 1.6
\enddata
\end{deluxetable}

We generally obtained 3 observations with integration times between 200 and 1200 seconds, depending on the set-up and the target brightness. The three observations are median-filtered to minimize contamination by cosmic rays.

We reduced the data using our spectroscopic data reduction pipeline. We subtract the bias and trim the overscan then flatten the image using dome flats. The spectra are extracted by fitting a Gaussian plus a linear background at each column.
The extracted spectrum is the area of the Gaussian fit at each wavelength. Uncertainties are based on counting statistics, including uncertainties in the fit background level. Wavelength calibration utilizes an arc lamp spectrum obtained before each set of images.

We obtained a spectrum of a spectro-photometric standard, generally either Feige~110 or LTT~4364, to provide a counts-to-flux conversion factor each night. Because this is slit spectroscopy, slit losses preclude an absolute flux calibration, but we do recover the shape of the spectrum. Small errors in the calibration arise from the fact that the target and the standard are generally observed at different air masses and consequent slit losses may change, and the spectroscopic slit is oriented E-W, not at the parallactic angle.

We calibrated the absolute fluxes of the spectra using the spectra obtained with the low dispersion grating 13 set-up, which spans the entire optical range including the $B, V, R_{\mathrm{C}}, $ and $ I_{\mathrm{C}}$ passbands. However, we only obtained four such spectra. Additionally, we were able to calibrate the absolute fluxes of the blue (grating 26 set-up) spectra in a two step process. The low dispersion blue spectra span the full width of the Johnson $B$ filter. We convolved these spectra with the filter response, and scaled the total flux to match the $B$ band flux interpolated to the time of observation. Correction factors ranged from 0.7 to 3, reflecting air mass corrections, slit losses, and changes in sky transparency between the time of these observations and those of our  spectral flux calibrator.

We then interpolated the continuum flux at $\lambda$~4250\AA\ for the calibrated low dispersion blue spectra to the times of observation of the high dispersion (grating 47/II) blue spectra, and scaled the latter accordingly.

\subsection{{\it Swift} observations}

The {\it Swift} satellite \citep{geh04} began observing Nova LMC 2009a on 2009 February 14, nine days after the optical discovery, with the last observation occurring almost a year later, on 2010 January 30. The resulting data were processed using the standard Swift tools within
HEASoft version 6.16, together with the most up-to-date calibration files.
Count rates were estimated using the XIMAGE package, while upper limits
were calculated using the Bayesian method of \cite{kra91}. Grades 0-12 were used for the Photon Counting (PC; time resolution of 2.51s) mode X-ray Telescope \citep[XRT;][]{bur05}  data, while grades $0-2$ were used for Windowed Timing data (WT; time resolution of 18~ms), which were collected when the source count rate was above about 1~count~s$^{-1}$. A circle with a 20 pixel (1 pixel = $2^{\prime\prime}\!\!.36$) radius centered on the source was used for the WT data, with the background determined from the same sized region offset from the source position. The PC data were considered to be piled-up above 0.4~count~s$^{-1}$, during which times annular extraction regions were used, excluding between two and seven pixels depending on the actual count rate. At other times, a 20~pixel radius circle was also used for the PC data, though this was decreased to 10 pixels when the source count rate was below $\sim0.1$~count~s$^{-1}$.

UV/Optical Telescope \citep [UVOT;] [] {rom05} source magnitudes were extracted within a $5^{\prime\prime}$ circular region centered on the source (since the aspect-correction failed for this field, the region was carefully re-centered on the source for every individual snapshot) and the background value calculated within a $15^{\prime\prime}$ radius offset from, but close to, the nova.

The UV source was detected throughout the {\it Swift} observations, with an X-ray source being detected from days 63 to 302. The latter in particular showed great variability and the cadence of observations was adjusted accordingly. Full details are given in Section 3.3 below.

\section{Results}

\subsection{SMARTS Photometry}\label{SMARTS_photom}

After \cite{lil09} reported the discovery of the nova on 2009 February 05.067 with an unfiltered TechPan film magnitude = 10.6, he found the nova had faded to magnitude 11.8 at 2009 February 07.092 (JD 4869.592\footnote{All JDs quoted here are less 2450000.0.}, $t = 2$ days). Our first observation, through an $I$-band filter only, occurred 2 minutes earlier than this latter observation, with $I=11.55$.  We make no attempt to convert Liller's magnitudes to $BVR_{\mathrm{C}}I_{\mathrm{C}}$.

\begin{figure}
\includegraphics[angle=0,width=\columnwidth]{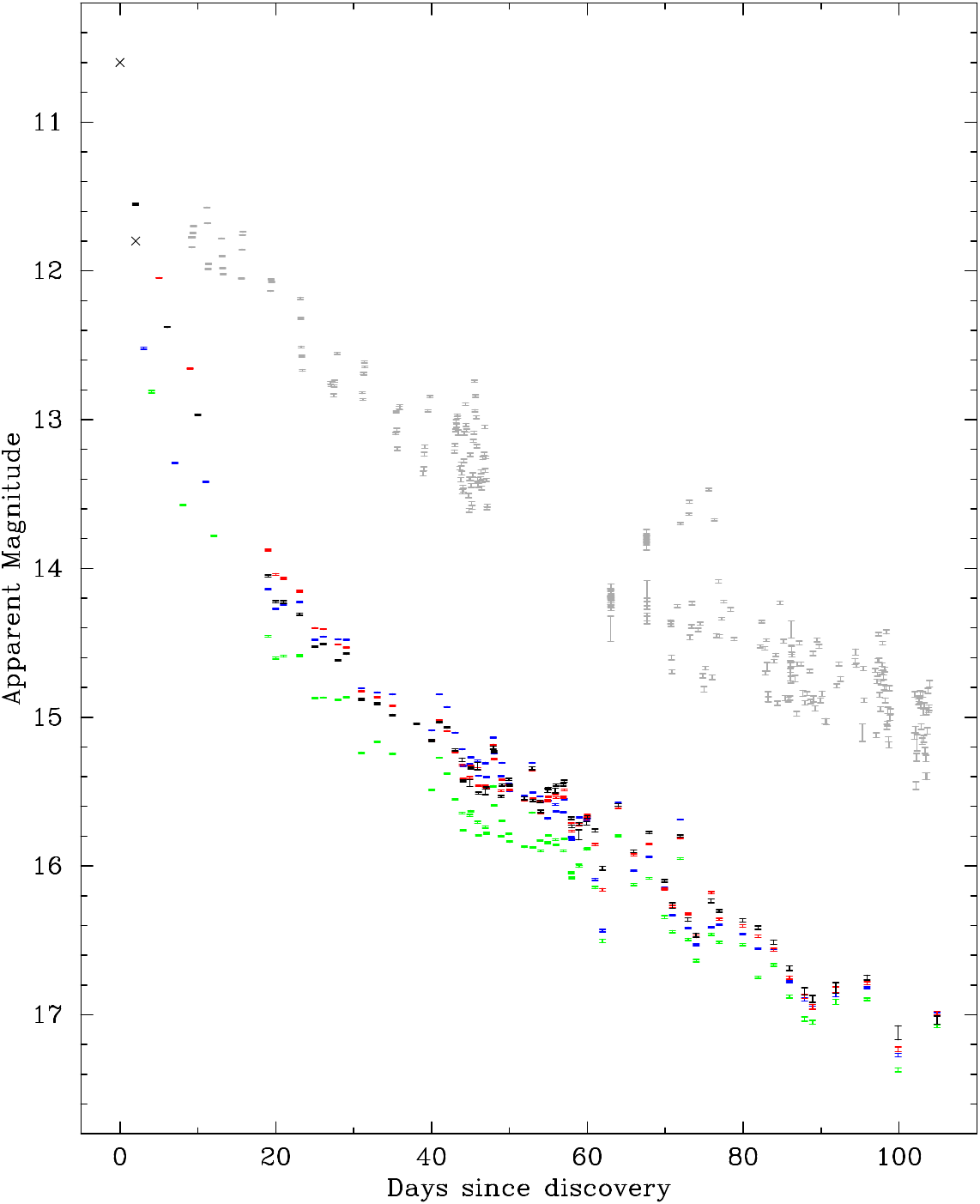}
\caption{The optical and near-UV (NUV) light curves of Nova LMC 2009a during the first 112 days post-discovery. Aside from an increase in the scatter between days 40 and 60, which may be related to the onset of the SSS emission and the 1.2 day periodicity (see e.g. Sections 3.3 and 3.4, respectively), the decay is monotonic.  The first black $\times$ indicates the (unfiltered) discovery magnitude \citep{lil09}, the second $\times$ an observation $\sim$2~days later.  The blue, green, red, and black data points show the SMARTS $BVRI$ photometry, respectively.  The SMARTS NIR $JHK$ photometry generally follows the optical light curve in this time-frame, but the larger uncertainties precluded plotting these data here. The gray data points indicate the {\it Swift}/UVOT uvw2 photometry.\label{fig1}}
\end{figure}

\begin{figure*}
\begin{minipage}[t]{0.49\textwidth}
\includegraphics[width=\columnwidth]{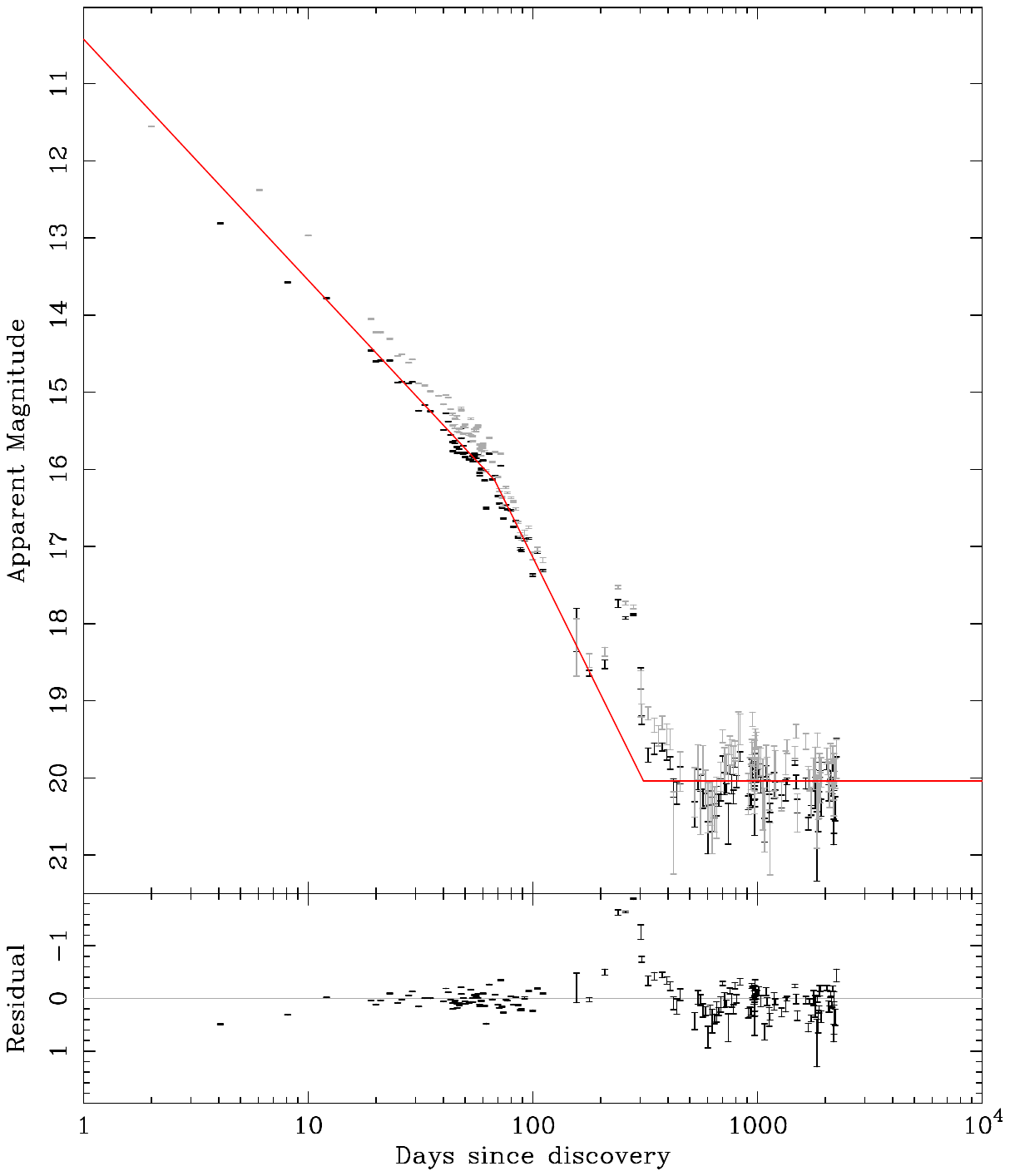}
\caption{Upper panel: The SMARTS $V$- (black) and $I$-band (gray) light curves of Nova LMC 2009a from 2009 February 6 through 2015 April 2 (days 1 to 2247 post-discovery, with the former also assumed to be the time of maximum light). The gaps around days 140, 500, and 860 are due to the nova position being close to the Sun.  The red line indicates an exponential fit to the $V$-band data between days 10 and 60, a break at day $\sim68$, a steeper exponential fit between days 70 and 200, and finally a fit to the quiescent data from day 400 onwards.  Lower panel: The residuals remaining following a subtraction of the `red' fit from the $V$-band data.  The gray horizontal line indicates the zero level.  We note here the apparent re-brightening between days $\sim200$ and $\sim400$ which is discussed further in Section~4. \label{fig2}}
\end{minipage}\hfill
\begin{minipage}[t]{0.49\textwidth}
\includegraphics[angle=0,width=\columnwidth]{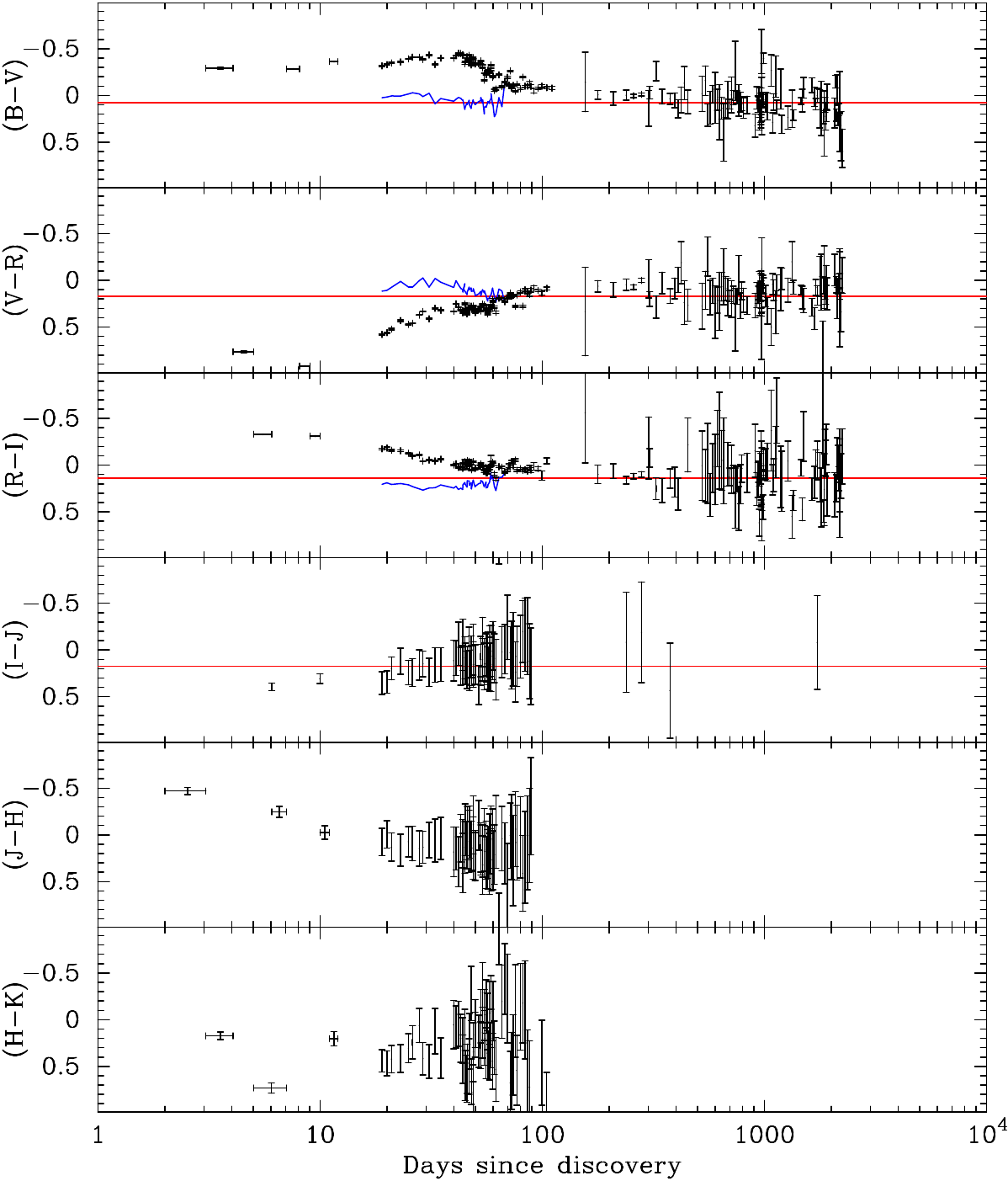}
\caption{The optical/NIR broadband color evolution of Nova LMC 2009a from day 1 to 2247 post-discovery.  The red lines indicate the mean color from day 400 onwards (where data exist).  The blue lines show the continuum color corrected for the line emission, as described in the text.  The $B-V$ color is bluest near day 40, after which it reddens. The $V-R$ color monotonically becomes bluer, while $R-I$ becomes redder. This reflects the decline in the emission line flux.\label{fig-cols}}
\end{minipage}
\end{figure*}

For the first four months of observations, the nova decayed monotonically, with no evidence for any re-brightenings or dust formation-related dimmings (see Fig.~\ref{fig1}). The $V$- and $I$-band light curves over the whole range of observations into the quiescent phase are shown in Fig.~\ref{fig2}. While we can fit our early light curve data fairly well with a single exponential decay through to day $\sim60$, that simple decay does not fit the first few data points, and the extrapolation of the exponential towards time $t=0$ obviously becomes meaningless. It is clear however that there is some indication of a trend towards Liller's discovery magnitude in the earliest data points.
Data from day $\sim70$ through day 200 can be well fit with a second exponential, and data from day 400 and 800 onwards, for the optical and NIR data, respectively, (assumed to be quiescence) can be fit with a constant magnitude.  An example of such a fit to the $V$-band data can be seen in Fig.~\ref{fig2}, with the residual from this fit shown in the lower panel.  A clear break in all the optical and NIR lightcurves is seen at day $65\pm2$, which coincides with the first detection of X-rays from the eruption \citep[][see Section~\ref{swift}]{bod09}.  From this point onwards, the emission begins to show more scatter.

In Table~\ref{tbl-phdecay} we provide the $t_2$ and $t_3$ timescales (days to decay 2 or 3 magnitudes from peak respectively). To determine these timescales we assume a peak magnitude of 10.6 in all bands (we note however the caveat that there is a possibility that the exact peak had been missed and the nova might have been even brighter at peak). The measurements of $t_{2}$ and $t_{3}$ were determined using an exponential fit with a timescale of the form $e^{-A+Bt}$, which is necessary to get close to the peak magnitude. A second (empirical) pair of $t_{2}$ and $t_{3}$ values were  generated from a linear interpolation between points and gave very similar results. 
The $t_2$ and $t_3$ timescales generally increase with wavelength, therefore the nova colors must initially get redder with time. By the time we have good color data, after day 2, the observed colors are becoming bluer. This may indicate that the peak $B$ and $V$ magnitudes did not reach 10.6, or that the evolution of the emission lines, generally in $B$ and $R$, is important. We come back to the emission lines later.

\begin{deluxetable}{lll}
\tablewidth{\columnwidth}
\tablewidth{0pt}
\tablecaption{Summary of  light curve decay timescales in different bands.\label{tbl-phdecay} }
\tablehead{
\colhead{Filter} & \colhead{$t_{2}$} & \colhead{$t_{3}$}}
\startdata
$B$  &    6.8 & 14.0 \\ 
$V$  &    5.0 &  10.4  \\
$R$  &    7.9 & 15.2 \\
$I$  &    7.2 & 14.2 \\
$J$  &   9.4 & 17.3 \\
$H$  &    9.9 & 18.4  \\
$K$  &   12.8 & 22.7 
\enddata
\end{deluxetable}

The light curve became more complex after the system returned from behind the Sun after $\sim 150$ days. Rather than the expected plateau while the soft X-ray source was bright \citep{kat08}, the brightness continued to decline until about day 200, after which it re-brightened by about two magnitudes in all bands (see Fig.~\ref{fig2}; this apparent `re-brightening' is further discussed in Section~\ref{rebrightening}). Following the turn-off of the X-ray source (around day 250--300; see Section~\ref{swift}) the system began to fade towards quiescence. The optical fading ceased after around day 400; we take the mean brightness after this date to be the quiescent level. Quiescent magnitudes obtained post-eruption by SMARTS are presented in Table~\ref{progen_phot}. 
Note that the system had faded below detectability in the NIR channels by day 110. 
We discuss the quiescent (progenitor) system in more detail in Section~\ref{quiescence}.

\begin{deluxetable}{llll}
\tablewidth{\columnwidth}
\tablewidth{0pt}
\tablecaption{Optical and NIR photometry of the progenitor system of Nova LMC 2009a.\label{progen_phot}}
\tablehead{
\colhead{Filter} & \colhead{Photometry} & \colhead{Source} & \colhead{Date}
}
\startdata
$B$  & $20.10\pm0.03$ & SMARTS & Mar 2010 -- Apr 2015 \\ 
$V$  & $20.04\pm0.02$ & SMARTS & Mar 2010 -- Apr 2015 \\
$R$  & $19.82\pm0.04$ & SMARTS & Mar 2010 -- Apr 2015 \\
$I$  & $19.75\pm0.03$ & SMARTS & Mar 2010 -- Apr 2015 \\
$J$ & $19.74\pm0.09$ & SMARTS & Apr 2011 -- Apr 2015 \\
$H$ & $19.2\pm0.1$ & SMARTS & Apr 2011 -- Apr 2015 \\
 & \\
$B$ & $18.32\pm0.37$ & ESO-B & Jan 1991 \\
$R$ & $19.20\pm0.62$ & ESO-R & Feb 1986  \\
 & \\
$B_{\mathrm{J}}$ & $18.17\pm0.20$ & SERC-J & Jan 1975 \\
  & \\
$J$ & $>19.3$ & 2MASS & Mar 1998 \\
$H$ & $>18.2$ & 2MASS & Mar 1998 \\
$K_{\mathrm{S}}$ & $>18.7$ & 2MASS & Mar 1998
 \enddata
\end{deluxetable}

We show the early color evolution in Fig.~\ref{fig-cols}. The broadband colors reflect the evolution of both the continuum and the emission lines. The color evolution through the first 120 days is essentially monotonic with some  variations superimposed, except in $B-V$, which becomes bluer until about day 40, and then becomes redder. 

We measured the fraction of the flux in the emission lines directly using our spectra (Table~\ref{tbl-linfrac}, where there are four nights where we obtained low dispersion grating 13 spectra covering the full optical range). 
The grating 26/Ia spectra span the entire $B$ filter, but other spectral set-ups  do not sample full broadband filter bandpasses, so we ignore them. We fitted a smooth continuum by eye between the emission lines. It is hard to define the continuum shortward of 4000\AA\ at the lowest resolutions, so  the uncertainty in the $B$ band line fractions can be up to 10~percent in those spectra. Emission lines are most important in $B$, with the lines contributing over $1/3$ of the flux until about day 50. The H$\alpha$ line is an important contributor to the $R$~band flux. Correcting the optical colors for the line contribution alters the color trends, and shows that the strongest contributions to the color changes are  the strong line emission in the $B$ band and the H$\alpha$ emission in the $R$ band (see Fig.~\ref{fig-cols}).

\begin{deluxetable}{lllll}
\tablewidth{\columnwidth}
\tablewidth{0pt}
\tablecaption{Fractional broadband flux in lines through given filters\label{tbl-linfrac}}
\tablehead{
\colhead{$t-t_{0}$ (days)} & \colhead{$B$} & \colhead{$V$} & \colhead{$R_{\mathrm{C}}$} & \colhead{$I_{\mathrm{C}}$}
}
\startdata
   7.068  & 0.31 & 0.11 & 0.49 & 0.21\\
  10.114  & 0.27 & \nodata  & \nodata & \nodata\\
  14.963  & 0.31 & \nodata  & \nodata & \nodata\\
  20.940  & 0.29 & \nodata  & \nodata & \nodata\\
  25.944  & 0.35 & \nodata  & \nodata & \nodata\\
  30.930  & 0.37 & \nodata  & \nodata & \nodata\\
  33.944  & 0.34 & \nodata  & \nodata & \nodata\\
  35.928  & 0.36 & \nodata  & \nodata & \nodata\\
  36.975  & 0.35 & \nodata  & \nodata & \nodata\\
  42.920  & 0.35 & \nodata  & \nodata & \nodata\\
  43.921  & 0.38 & \nodata  & \nodata & \nodata\\
  45.965  & 0.41 & 0.07 & 0.23 & 0.06\\
  50.920  & 0.37 & \nodata  & \nodata & \nodata\\
  52.926  & 0.31 & \nodata  & \nodata & \nodata\\
  54.914  & 0.23 & \nodata  & \nodata & \nodata\\
  57.918  & 0.26 & \nodata  & \nodata & \nodata\\
  64.911  & 0.18 & \nodata  & \nodata & \nodata\\
  65.914  & 0.24 & 0.07 & 0.14 & 0.05\\
  66.908  & 0.14 & 0.05 & 0.13 & 0.05\\
  68.934  & 0.16 & \nodata  & \nodata & \nodata\\
  72.908  & 0.12 & \nodata  & \nodata & \nodata\\
  84.905  & 0.10 & \nodata  & \nodata & \nodata\\
  96.894  & 0.09 & \nodata  & \nodata & \nodata
\enddata
\end{deluxetable}

\begin{figure}
\includegraphics[width=\columnwidth]{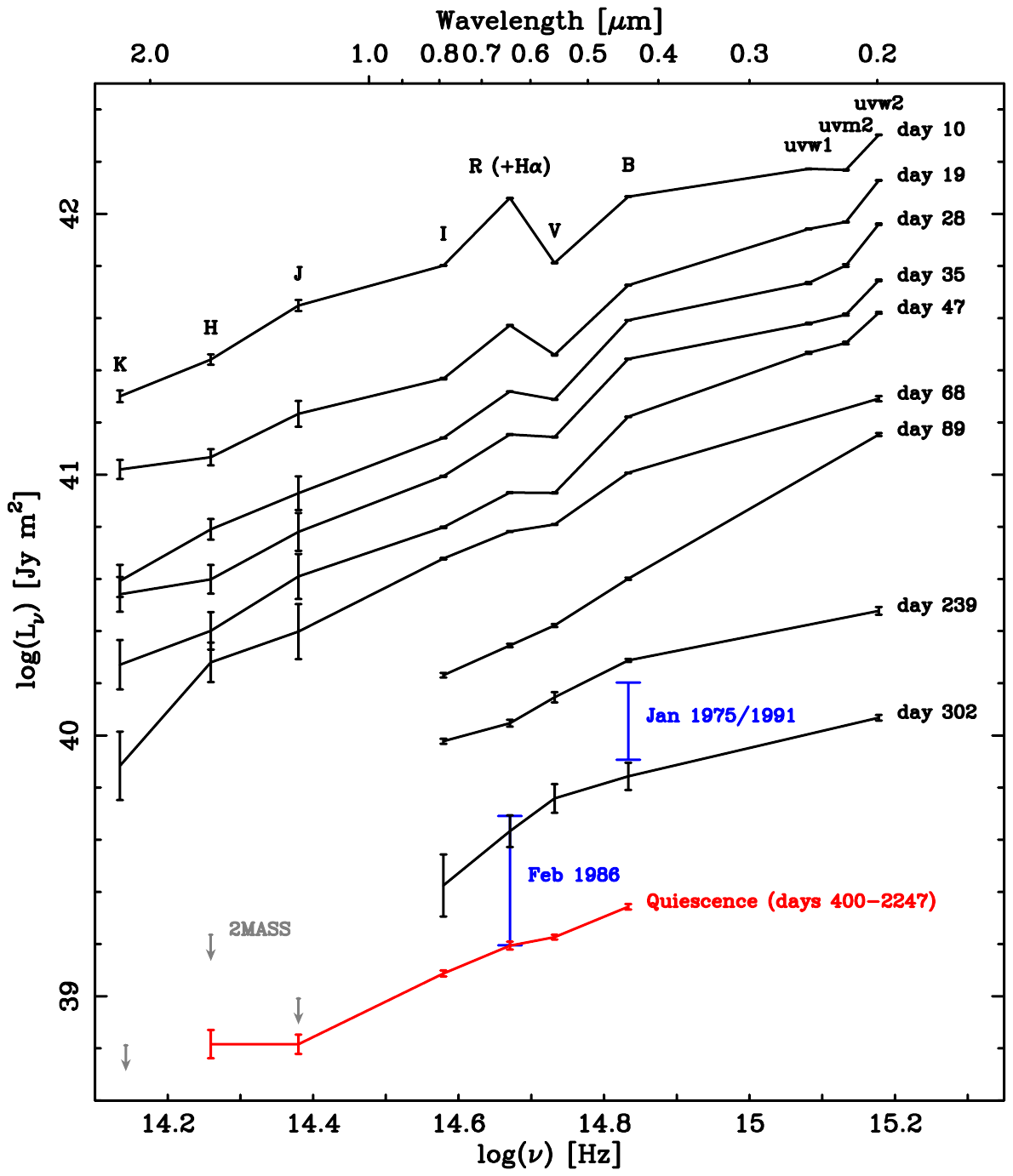}
\caption{Distance and extinction corrected SED showing the evolution of Nova LMC 2009a from day 10 post-discovery through to quiescence (red data points).  The blue data points indicate two independent inter-eruption detections (see Section~\ref{quiescence}).  The gray points indicate three inter-eruption upper limits obtained from re-analysis of the 2MASS data.  Here only the photometric uncertainties are shown (the magnitude of systematic uncertainties from distance and extinction are shown in the similar Fig.~\ref{fig19}).  The central wavelength locations of the optical/NIR $BVRIJHK$ filters and the NUV {\it Swift}/UVOT uvw1, uvm2, and uvw2 filters are shown to assist the reader. \label{fig4}}
\end{figure}

The evolution of the broadband photometric data from the eruption and subsequent decline of Nova LMC 2009a are shown in the distance and extinction corrected spectral energy distribution (SED) presented in Fig.~\ref{fig4}.  We have followed the methodology of \citet{schae10} to enable comparison with their Galactic RN SEDs (see their Fig.~71).  For Nova LMC 2009a we assume a distance to the LMC of $d = 48.1\pm2.3_{\mathrm{r}}\pm2.9_{\mathrm{s}}$ kpc \citep{mac06}
and an estimated extinction towards the LMC of $A_{V}=0.6\pm0.2$ (see Section~\ref{disc}).  In this figure, SEDs are shown for days 10, 19, 28, 35, 47, 68, 90, 239, 302, and finally an average over day 400--2247 (in red), which is assumed to be quiescence.  A number of inter-eruption data points are also shown -- the gray points indicate upper limits from 2MASS data, and the blue points optical detections (see Section~\ref{quiescence} for further details).

The optical behaviour of novae around peak has been observed to resemble black body emission \citep{gal76,geh08} and has been modeled in terms of the development of a pseudo-photosphere (PP) in the optically thick ejecta \citep[see][and references therein]{bat89}. The PP radius $r_{\rm p}$ is greatest, and effective temperature $T_{\rm p}$ at a minimum at optical peak. Thereafter, the mass loss rate from the WD surface declines, and at constant bolometric luminosity, $r_{\rm p}$ shrinks while $T_{\rm p}$ rises, shifting the peak of the emission further into the UV with time past maximum light. In CNe, with relatively high ejected masses, $T_{\rm p} \sim10,000$ K at optical maximum placing the peak emission in the optical part of the spectrum. 

Fig.~\ref{fig4} shows that if the simple PP model is correct, then the peak of the continuum emission throughout the epochs sampled is at wavelengths shorter than that of the uvw2 filter on {\em Swift} ($1928$\AA\ ; i.e. $T_{\rm p} >15,000$K). For this very fast nova, with first SED observations taken 10 days after peak, this is not unexpected (for a more detailed treatment of the evolving nova spectrum at outburst, see \citealt{hau08}). 

\subsection{SMARTS Spectroscopy}

\begin{figure*}
\includegraphics[width=\textwidth]{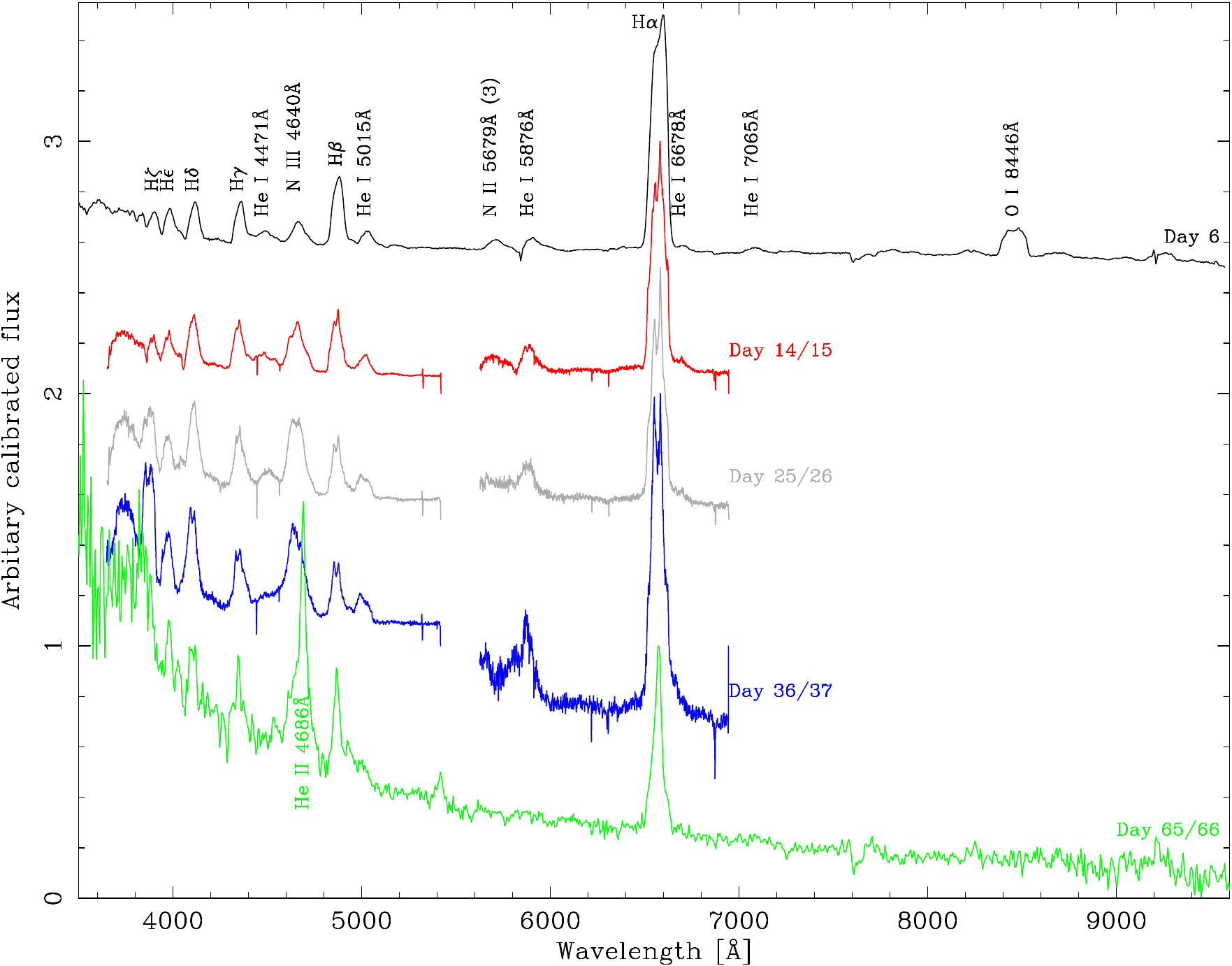}
\caption{Spectra of Nova LMC 2009a taken with the RC spectrograph on the SMARTS 1.5m telescope on day 6 (black), 14/15 (red), 25/26 (gray), 36/37 (blue), and 65/66 (green), see text for full details.  Selected emission lines have been labelled at their LMC rest wavelengths.
\label{fig5}}
\end{figure*}

\begin{figure}
\includegraphics[width=\columnwidth]{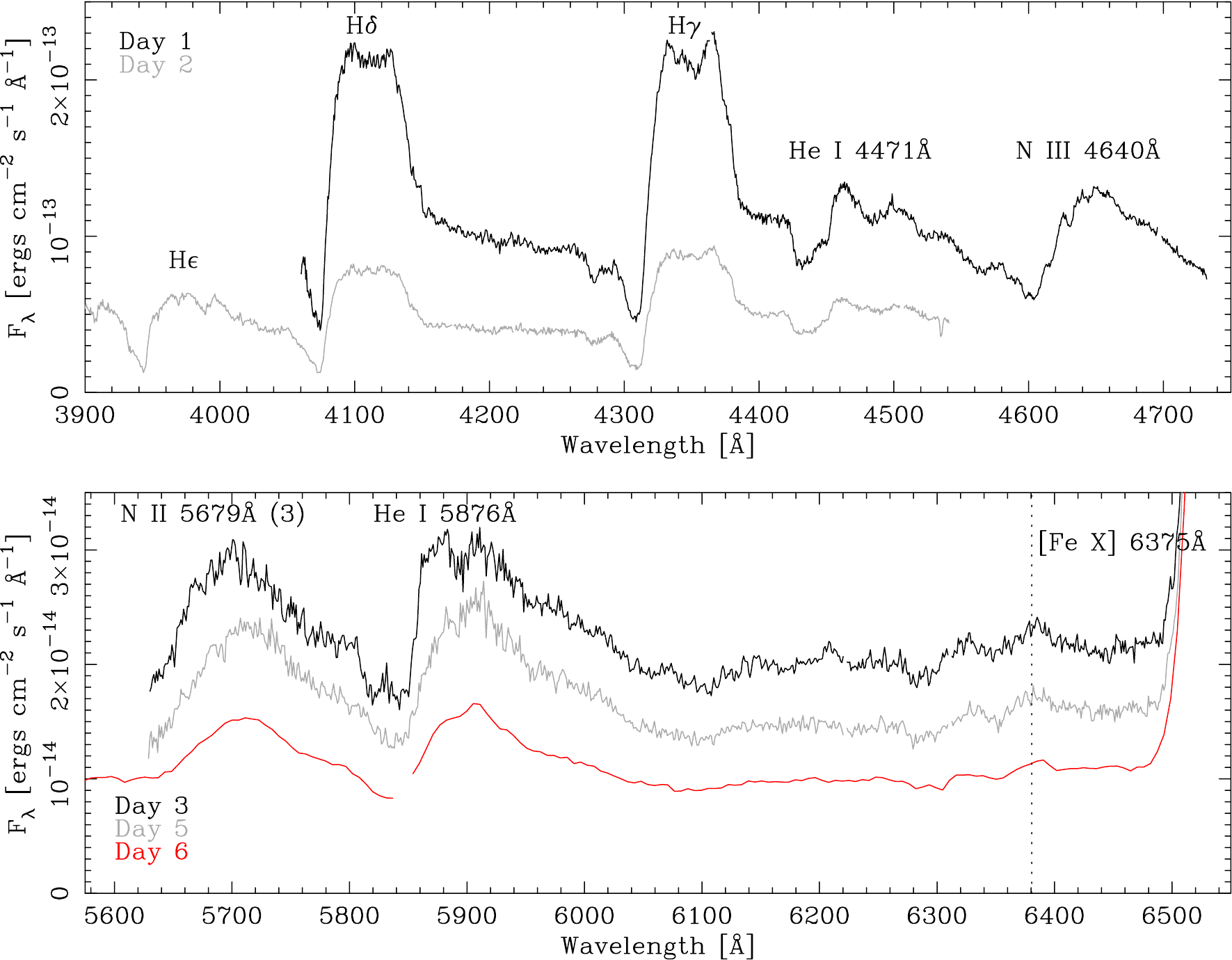}
\caption{Flux calibrated spectra of Nova LMC 2009a taken with the RC spectrograph on the SMARTS 1.5m telescope, with selected emission lines labelled.  Top: Spectra taken on day 1 (black) and day 2 (gray) post-eruption.  Here, P Cygni absorption profiles are clearly visible around the Balmer lines.  Bottom: Spectra taken on day 3 (black), day 5 (gray), and day 6 (red; see also black line in Fig.~\ref{fig5}).  Here, P Cygni profiles are no longer observed and the location of a faint highly ionised [Fe~{~\sc x}] emission line is indicated (see text for further discussion). \label{fig6}}
\end{figure}

In Fig.~\ref{fig5} we present five low/medium resolution SMARTS spectra of Nova LMC 2009a, taken on day 6 (black spectrum; set-up 13/I; see Table~\ref{tbl-spsetups}), 14/15 (red; set-ups 26/Ia and 47/Ib, respectively), 25/26 (gray; set-ups 26/Ia and 47/Ib, respectively), 36/37 (blue; set-ups 26/Ia and 47/Ib, respectively), and 65/66 (green; co-added, both set-up 13/I).  All spectra are dominated by broad emission lines of the Balmer series, as well as emission from He and N, confirming the classification of LMC 2009a as a He/N nova in eruption.  The early spectra show a flat continuum that evolves towards a bluer continuum at later times (see e.g. Fig.~\ref{fig5}). Figure~\ref{fig6} shows some of the early spectral evolution in more detail, including that of the observed P~Cygni absorptions. 

\subsubsection{He~{\sc ii} $\lambda$4686}

Because the first ionization potential of He is about 50~eV, the detection of He~{\sc ii} $\lambda$4686 indicates the presence of high energy photons or particle collisions. He~{\sc ii} $\lambda$4686 lies on the side of the strong $\lambda\lambda4640-4650$\AA\ Bowen fluorescence blend, the strength and shape of which varies with time. We see no evidence of any excess emission at $\lambda$4686\AA\ prior to day 29.

\begin{figure}
\includegraphics[angle=0,width=\columnwidth]{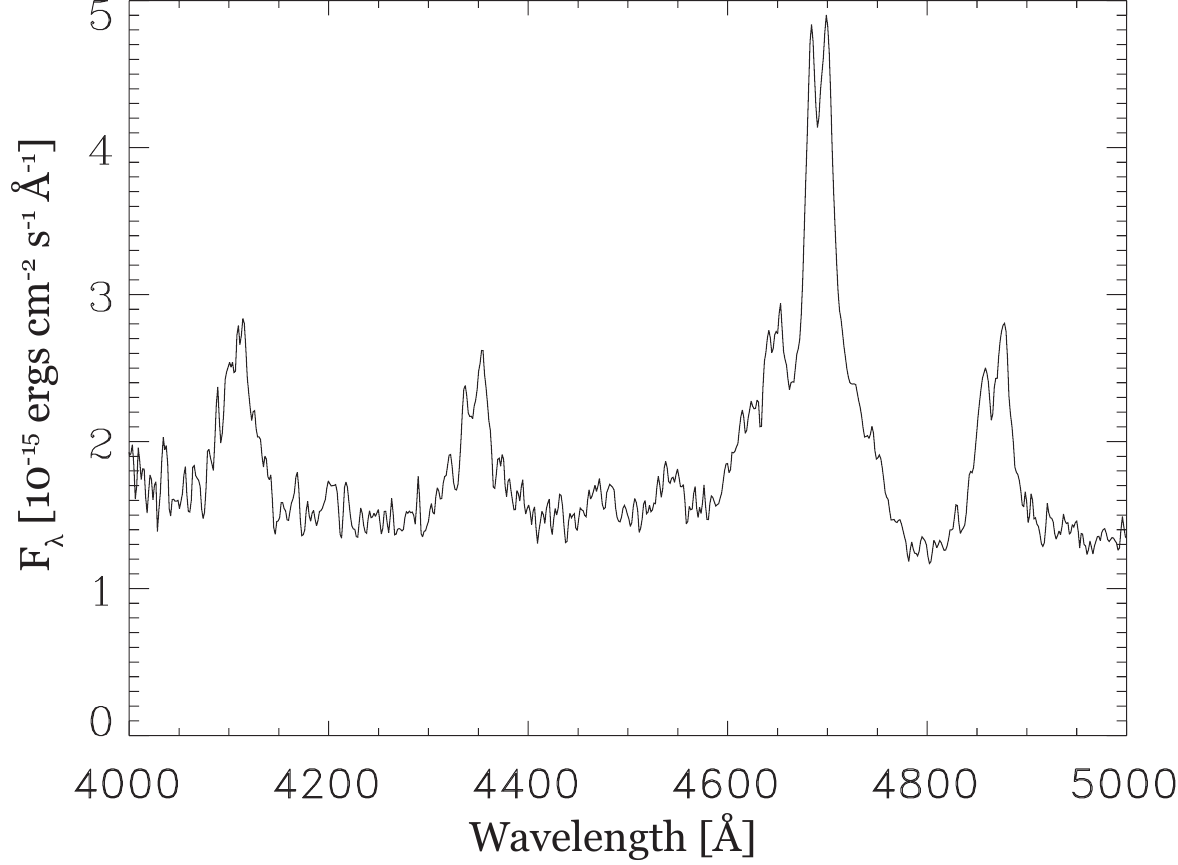}
\caption{Spectrum taken with the RC spectrograph on the SMARTS 1.5m telescope on day 54.9. This is the result of a $3\times600$ second exposure and has been smoothed with a Fourier filter. The strongest line is He~{\sc ii} $\lambda$4686. It is double-peaked and sits on the side of the broad 4640~\AA\ Bowen blend. Also visible are three Balmer lines, H$\beta$, H$\gamma$ and H$\delta$, which are also double-peaked. Other lines possibly present are He~{\sc ii} $\lambda$4200, He~{\sc i} $\lambda$4471, He~{\sc ii} $\lambda$4542.\label{fig7}}
\end{figure}

\begin{figure}
\includegraphics[angle=0,width=\columnwidth]{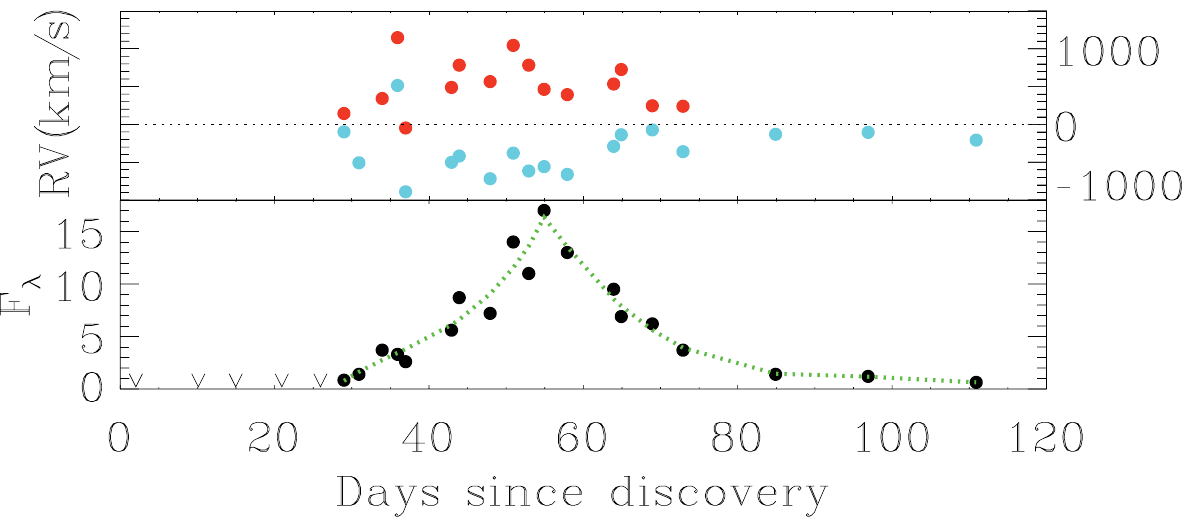}
\caption{The evolution of the velocities of the components (top panel) and the flux (in units of $10^{-14}$ erg cm$^{-2}$ s$^{-1}$; bottom) in the He~{\sc ii} $\lambda$4686 line as a function of days since discovery. The radial velocities are of the two components to the line, fit as Gaussians, relative to the rest wavelength of the LMC (dotted line). The lines appear to separate until about day 50, after which they start to converge again. Only a single component is visible after day 80. Absolute fluxes are determined as stated in the text and RMS errors on the measured velocities are estimated as typically around $45$ km s$^{-1}$. No line is visible before day 29 and the flux peaks near day 55. The dotted green line is a third order polynomial independently fit to the rising and falling legs.  We attribute no physical significance to the functional form. \label{fig8}}
\end{figure}

A narrow emission feature above the local continuum (the red edge of the Bowen blend) becomes visible on day 29. Although it is difficult to discern in this earliest spectrum due to the relative dominance of the Bowen blend, the emission line is double-peaked (see e.g. Fig.~\ref{fig7}), with the mean velocity centered about the rest wavelength of He~{\sc ii} at the 278 km s$^{-1}$ redshift of the LMC. As shown in Fig.~\ref{fig8}, the emission flux rises rapidly until about day 55, whence it begins to decay approximately equally fast. Weak emission is detectable through our last observation on day 112.

We fit the emission line spectra as two Gaussian components atop a background parameterized as a third-order polynomial. The two components do not have the same width. Early on, the weak emission, in particular the bluer component, is difficult to discern against the steep background. We note that the existence of any spectral features in this region that we have not accounted for could bias the measurements.

The emission lines were double-peaked upon discovery. The velocity splitting increased to about $\pm800$ km/s and then decreased. Within the limits of our S/N we cannot quantify exactly when the trend reversed, but the velocity splitting  appears to have peaked before the flux peaked. 

 \subsubsection{H$\alpha$}

\begin{figure}
\includegraphics[angle=0,width=\columnwidth]{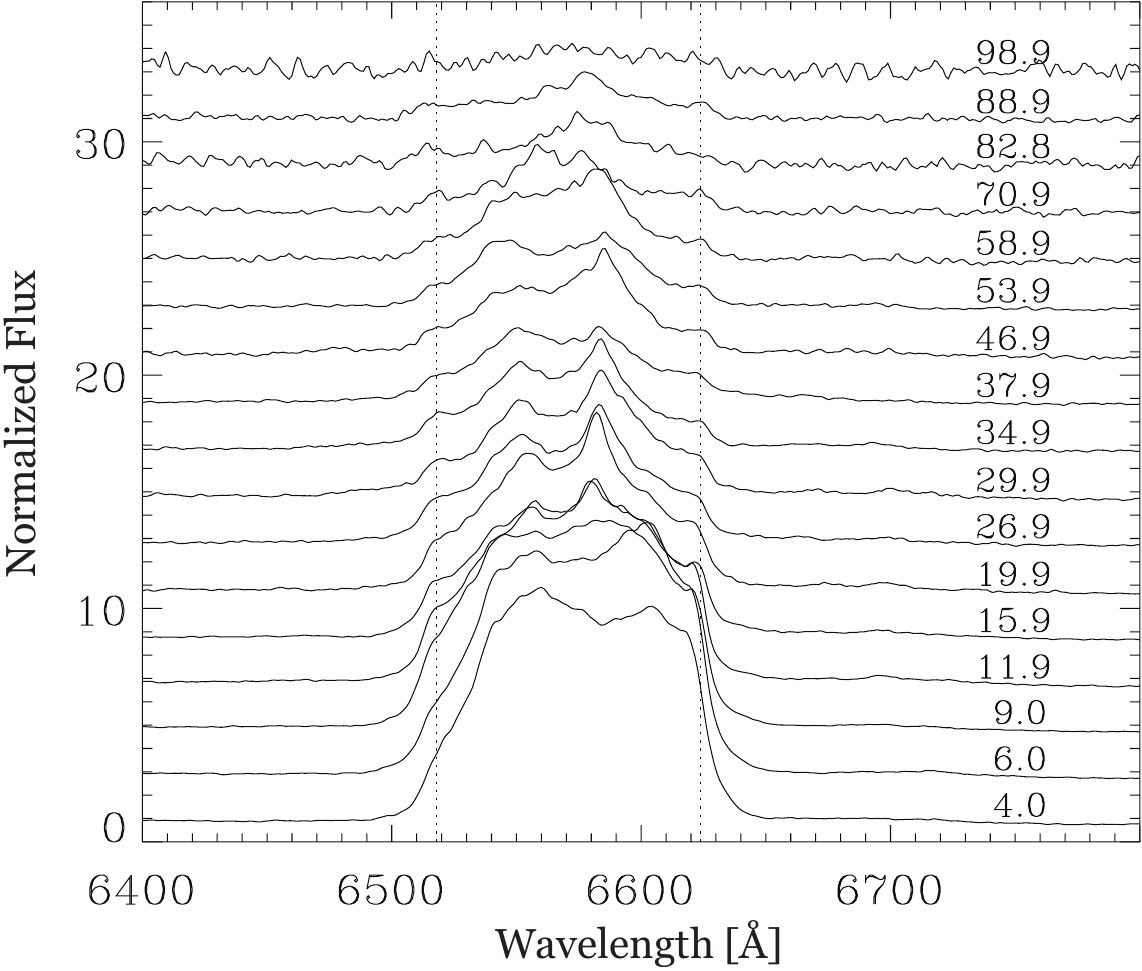}
\caption{Evolution of the H$\alpha$ line profile. Numbers on the right of each spectrum are days since $t = 0$. Spectra are normalised to the continuum (=1). The gray dotted lines mark the positions of the shoulders in each spectrum, with little evidence for any velocity evolution. The fluxes are plotted linearly with offsets of two units between spectra for clarity.\label{fig9}}
\end{figure}

\begin{figure}
\includegraphics[angle=0,width=\columnwidth]{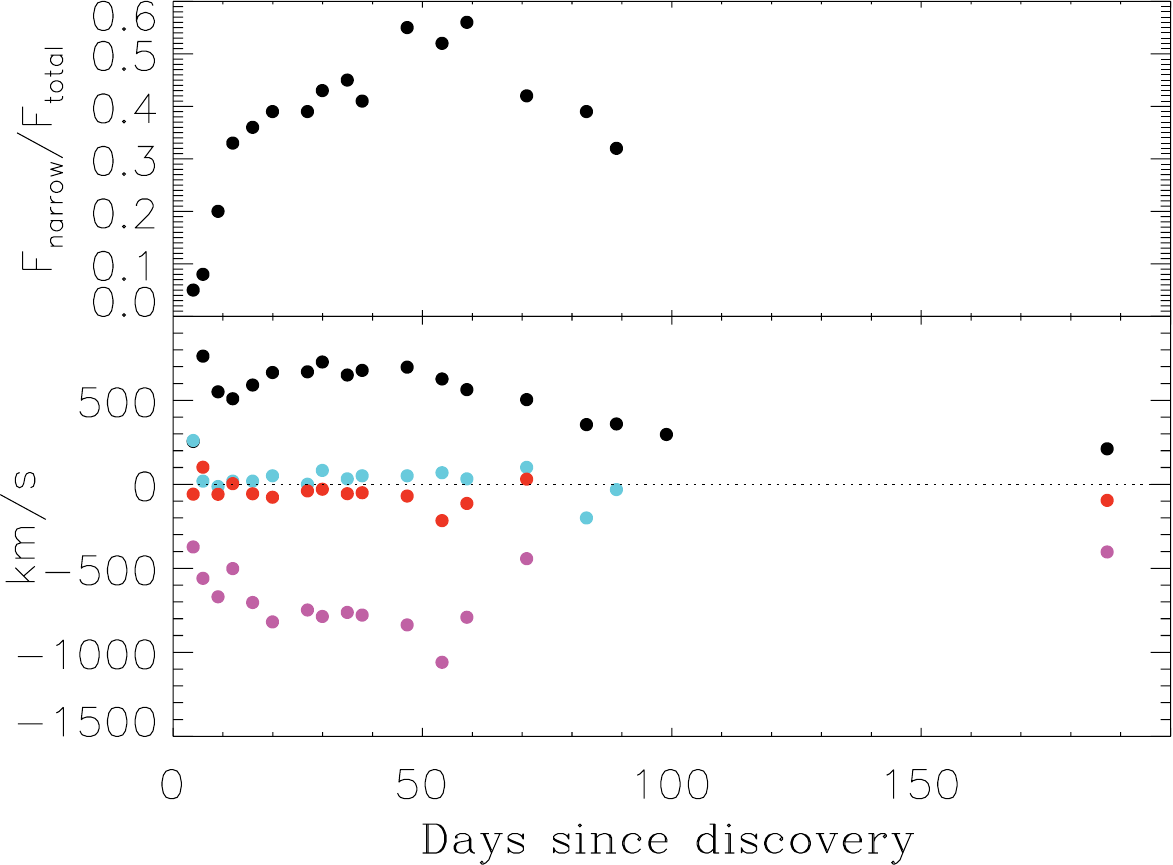}
\caption{Properties of the H$\alpha$ line. We deconstruct the line into a pedestal plus a narrower double-peaked central emission which we fit as the sum of two Gaussians. The line profile is more complex, and this simple fit is generally poor, but we can accurately determine the wavelengths of the  two peaks. The top panel shows the fraction of the H$\alpha$ flux in the narrower central component. This fraction peaks around the time of the He~{\sc ii} $\lambda$4686 flux maximum and when the X-rays were first detected. The lower panel shows the velocities of the dominant Gaussian components (black and magenta) with respect to the mean LMC velocity. The mean of these two velocities is shown in red. Within the uncertainties, it  agrees with the rest velocity of the LMC. The aqua points are the mean velocities of the broader pedestal. Estimated RMS errors on velocities are less than the 45 km s$^{-1}$ found for the He~{\sc ii} lines as given in Fig~\ref{fig8}.\label{fig10}}
\end{figure}

The H$\alpha$ line exhibits a complex and evolving profile (Fig.~\ref{fig9}).  Early on the emission is dominated by a broad ($\pm\sim$2500 km/s)
component that rapidly fades while maintaining its full velocity width (the 
``shoulders'' in Fig.~\ref{fig9}). The two stationary high velocity peaks are commonly
seem in recurrent novae \citep[e.g.][]{wal11}.
A narrower double-peaked central component becomes prominent after a few days,
and soon dominates the flux in the line (Fig.~\ref{fig10}).

We fit the central component as the sum of two Gaussians superposed on the pedestal. It is clear that the profile is more complex, but this suffices to quantify the major structures in the line. Some of these are shown in Fig.~\ref{fig10}. The emission centroids of the two components behave like those of He~{\sc ii} $\lambda$4686, with a velocity separation that increased to about $\pm700$ km s$^{-1}$ by day 50 and decreased thereafter. The mean of the two velocities is $+34$ km s$^{-1}$ with respect to the LMC standard of rest. The mean velocity of the pedestal component, taken to be the mean of the FWHM velocities, is $-52$ km s$^{-1}$. Given  the uncertainties these are not significantly different.

\subsubsection{He~{\sc i}}

The He~{\sc i} lines at 5876\AA\ and 6678\AA\ are detected in the first red spectrum, four days after discovery. The lines are clearly double-peaked as early as day 10, but are relatively weak and blended - $\lambda$5876 with the Na~{\sc i} doublet in absorption and $\lambda$6678 with the bright wing of H$\alpha$. Qualitatively, the two lines behave the same as H$\alpha$ and He~{\sc ii} $\lambda$4686.

\subsubsection{Other Lines}

We have used the line lists in \cite{wil94} to guide our search for other atomic species in these spectra.

Line identification in the blue is difficult due to the extreme line widths. The upper Balmer lines have FWHMs from 35\AA\ (H$\epsilon$, $\equiv 2650$ km s$^{-1}$) through 63\AA\ (H$\beta$, $\equiv 3900$ km s$^{-1}$) on day 10. In addition to the Balmer lines through H$\zeta$ (H-8), the Bowen blend at $\lambda$4640\AA\ and [O~{\sc iii}]/He~{\sc i} $\lambda\lambda$5007/5015\AA\ are the strongest features. Line blending precludes identification of discrete features blue-ward of H$\zeta$ $\lambda$3888.

The red spectra are dominated by H$\alpha$ at all times. Early on, the next strongest lines are He~{\sc i} $\lambda$5876 and the $\lambda\lambda$5700 N~{\sc ii} blend. The 5700\AA\ blend is visible up to  day 20; the He~{\sc i} lines are visible up to day 59 but are not detected on day 71. Na~{\sc i} from the nova ejecta is detected against the background of the  He~{\sc i} $\lambda$5876 emission line up to day 16.

There is narrow emission coincident with  [Fe~{\sc x}] $\lambda$6375 on day 4 (see Fig.\ref{fig6}). The [Fe~{\sc x}] line is still visible on day 6, but is not discernible by day 9. There is a marginal detection of a narrow line at 6200\AA\ on days 2 and 4. This is coincident with a permitted O~{\sc vi} line. Overall, we see evidence for an early very high ionization phase that ended within 9 days. The presence of [Fe~{\sc x}] indicates electron temperatures of around $10^6$K \citep{gor72} and is associated with either ionization/excitation from shocks or a central Super-Soft Source \citep[][and see below]{sch11}. However, the SSS is neither expected to appear, nor is detected around this time (the first {\it Swift} observation being at day 9, see also next section), and a more likely origin is either intra-ejecta or ejecta-pre-existing circumstellar medium shocks. Early harder X-ray emission is seen in several Galactic novae observed by {\it Swift} \citep{sch11}. Such emission is particularly bright in RG-novae such as RS Oph, arising from shocks driven into the pre-existing red giant wind by the impacting ejecta \citep[e.g.][]{bod06}. For RS Oph at the distance of the LMC, we expect such hard X-ray emission to have peak XRT flux of $\sim 0.03$ cps, and thus have been detectable. However, for most novae (those without RG secondaries with extensive pre-eruption winds), although hard X-ray emission at early times is fairly common, and e.g. seen in U Sco \citep{sch11}, it is not expected to be detectable at the distance of the LMC. For example, the hard X-ray component seen until the emergence of the SSS in KT Eri around 65 days after eruption \citep{sch11} would have had an XRT flux of $\sim 7\times10^{-4}$ cps (see also Fig.\ref{fig22}). There is some evidence for the existence of a hard X-ray component in LMC 2009a at later times however (see Section 3.3 below).

P~Cygni absorption profiles are seen around the spectral lines, particularly H$\gamma$ and H$\delta$, in the high resolution blue spectra taken on day 1 and day 2, set-ups 47/IIb and 47/II, respectively (see Table~\ref{tbl-spsetups} and Fig.~\ref{fig6}).  No P~Cygni profiles were visible in the red spectrum taken on day 3, or in any subsequent spectra. Expansion velocities derived from the P~Cygni profiles are as follows: day 1 -- H$\gamma \sim3610$ km s$^{-1}$, H$\delta > 3180$ km s$^{-1}$; day 2 -- H$\gamma \sim3650$ km s$^{-1}$, H$\delta \sim3300$ km s$^{-1}$, H$\epsilon \sim4090$ km s$^{-1}$ with estimated errors of $\pm100$ km s$^{-1}$ on each (it is difficult to derive the velocities more accurately due to the flat-topped, $\sim 50$\AA-wide emission line profiles). 

\subsection{{\it Swift} Data}\label{swift}

\begin{figure*}
\includegraphics[angle=270,width=\textwidth]{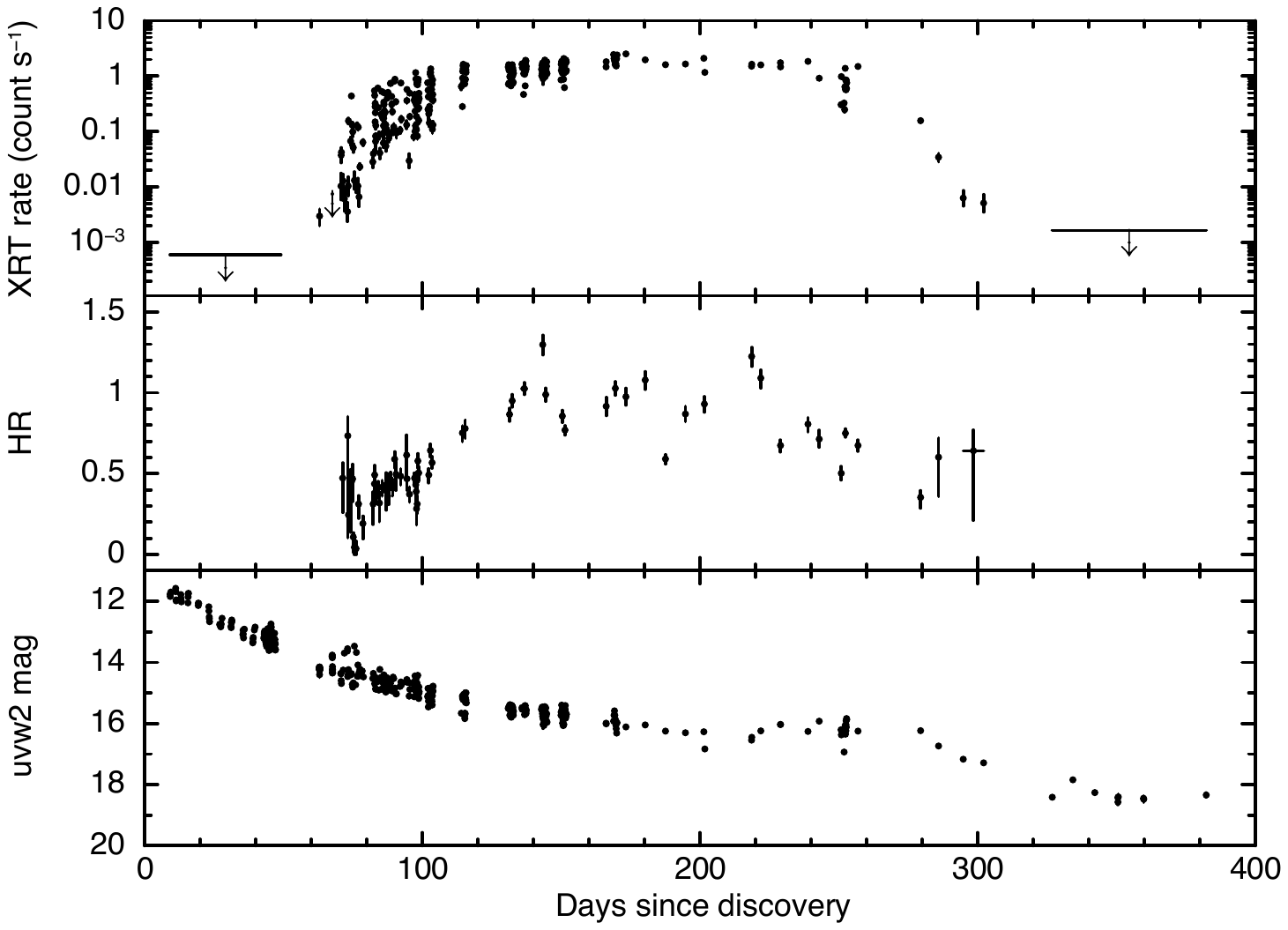}
\caption{Evolution of the {\it Swift} XRT 0.3-10 keV count rate (counts s$^{-1}$; top panel) and UVOT uvw2 magnitude (bottom panel). Also shown is the evolution of the X-ray hardness ratio (HR, [0.45-10]/[0.3-0.45] keV). Upper limits are at the $3\sigma$ level. Fig.~\ref{fig14} shows a section of the UVOT light curve observed at high cadence.\label{fig11}}
\end{figure*}

At the time of the first {\it Swift} observation at $t = 9$ days, a UV source was clearly detected in all three UVOT NUV filters (uvw1, uvm2 and uvw2, with central wavelengths of 2600, 2246 and 1928 \AA ~respectively), as first announced by \cite{bod09}. Observations continued every few days for the next two months, with the UV magnitude fading from about 11.5 to 14, but also showing short-term variability with peak-to-peak amplitude $\sim$0.6 mags (see Fig.~\ref{fig11}). High cadence observations (snapshots of data obtained almost every 96 minute spacecraft orbit) from days 43.0-46.9 after eruption demonstrated that the variability included a periodic component with $P \sim$ 1.2 days \citep[][see Section~\ref{oscillations}]{bod09}.

On day 63.0, there was the first marginal detection (99.7~percent confidence) of an X-ray source, at a count rate of $(3 \pm 1)\times10^{-3}$ count s$^{-1}$ \citep{bod09b}. From day 70, the X-ray source was clearly detected, with a strongly variable count rate observed; the emission was found to be dominated by a super-soft spectral component (see below).

Daily observations were performed between days 70-79 and again between days 82-90, at which point the cadence decreased slightly. Throughout this time, the UV source showed an overall fading trend, although with the periodic oscillations superimposed.

Further high-cadence observations between days 97-152 after eruption demonstrated that the X-ray source was also varying periodically (again, see further discussion in Section~\ref{oscillations}). 

Observations then continued approximately once a week for six months.  The UV source faded until around day 200, after which there was an apparent re-brightening (see Section~\ref{disc}), followed by a final decline. The maximum X-ray count-rate ($\sim$2.5 count s$^{-1}$) was detected on day 173, after which a fading trend (although still with the variations superimposed) began. Some time between day 257 and 279, when no data were collected, the count-rate began to fall away, until the final X-ray detection occurred 302 days after eruption. The UV source remained visible until the end of the observing campaign.

\begin{figure}
\includegraphics[angle=270,trim=70 10 40 80,clip,width=\columnwidth]{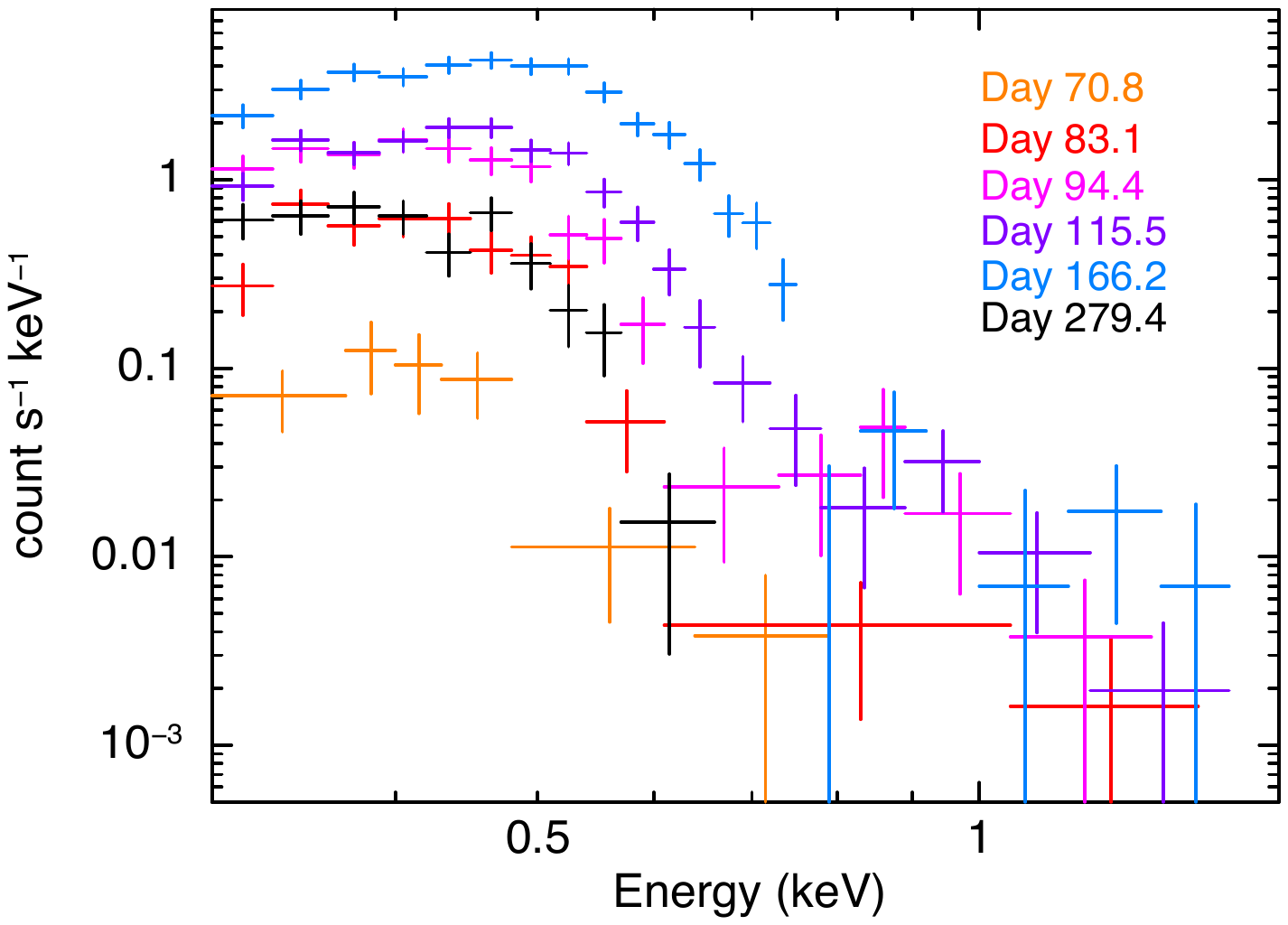}
\caption{Evolution of the observed {\it Swift} XRT spectrum. The FWHM of the spectral response kernel at the time of these observations was 0.11~keV at 0.5~keV in 2009.\label{fig12}}
\end{figure}

\begin{figure}
\includegraphics[angle=270,trim=70 10 40 80,clip,width=\columnwidth]{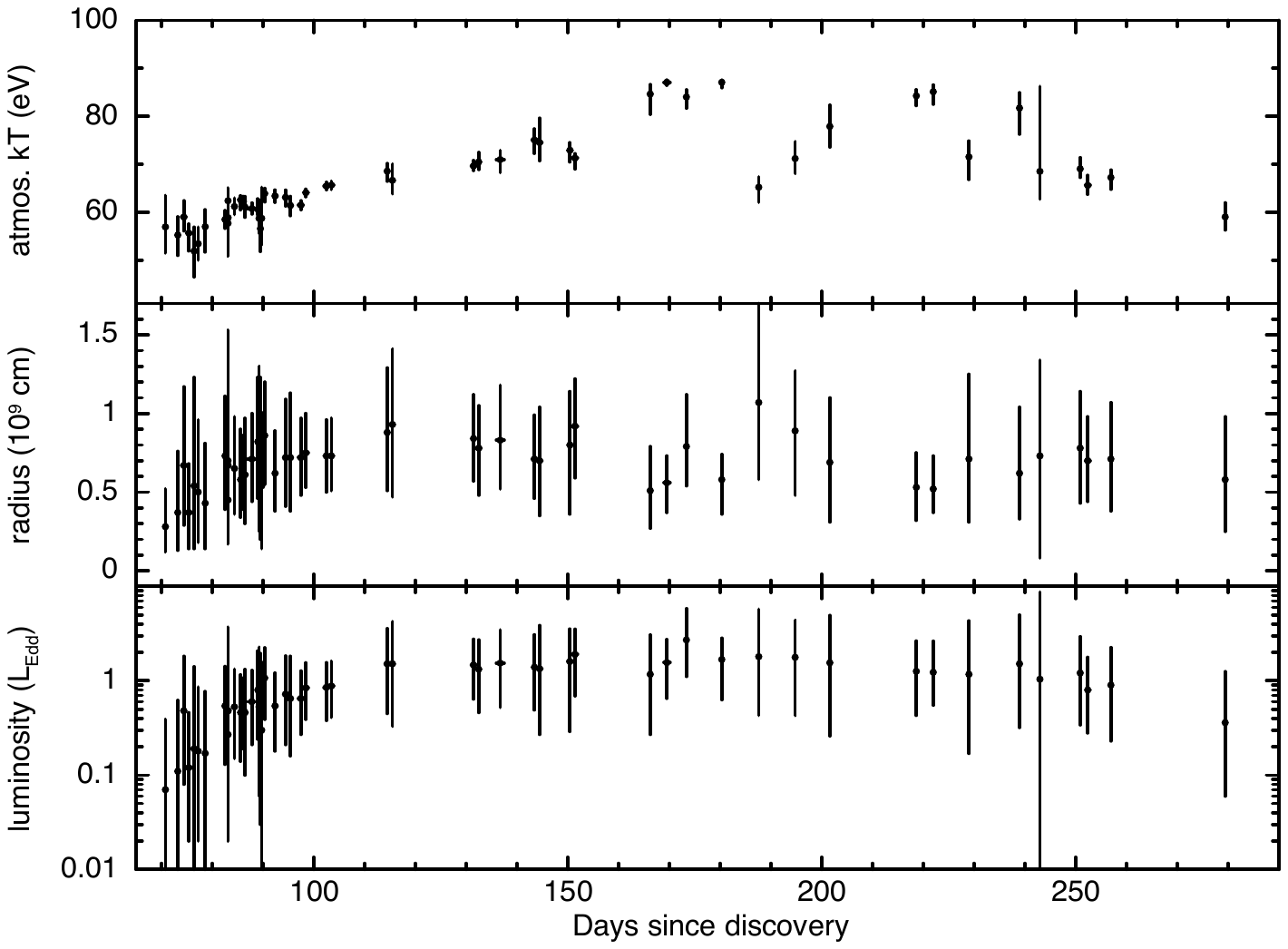}
\caption{Parameters derived from model atmosphere fits to the {\it Swift} XRT data (see text for details). Error bars indicate 90\% confidence limits. $L_{Edd}$ is the Eddington luminosity for a $ 1.2 {\rm M}_\odot$ white dwarf ($\sim 3.8\times10^4 {\rm L}_\odot$) \label{fig13}}
\end{figure}

Figure~\ref{fig12} shows the evolution of the observed X-ray spectrum. Figure \ref{fig13} then shows the results of fitting the XRT spectra with a plane-parallel, static, non-local thermal equilibrium (NLTE) atmosphere model (grid 003\footnote{\url{http://astro.uni-tuebingen.de/\textasciitilde rauch/TMAF/flux\_HHeCNO\\NeMgSiS\_gen.html}. In the framework of the Virtual Observatory ({\it VO}; \url{http://www.ivoa.net}), these spectral energy distributions (SEDs, $\lambda - F_\lambda$)  are available in {\it VO} compliant form via the {\it VO} service {\it TheoSSA} (\url{http://vo.ari.uni-heidelberg.de/ssatr-0.01/TrSpectra.jsp?}) provided by the {\it German Astrophysical Virtual Observatory} ({\it GAVO}; \url{http://www.g-vo.org}).; \cite{rau03}; \cite{rau10}}) and an optically thin plasma model (Mekal) component (with absorption fixed at $N_{\rm H} = 1.13\times10^{21}$cm$^{-2}$, derived from HI maps by \cite{kal05}, known as the Leiden/Argentine/Bonn (LAB) survey). Most of the spectra show at least a few counts above 1-2 keV. One spectrum was extracted per Obs ID (median duration 0.2 days), some of which span a significant variation in intensity and hardness ratio. Data taken after day 280 are not included because there are insufficient counts to usefully constrain spectral fit parameters during the decline phase. 

From the model atmosphere fits to the evolving XRT spectrum, the temperature, radius and luminosity were determined for
the SSS (see Fig.~\ref{fig13}); the normalisation of atmosphere models is defined in terms of the effective emitting radius and distance to the object. A slow rise in temperature to a peak of k$T = 87$ eV around day 170-180, followed by an unexpectedly noisy decline is evident. The measured radius and luminosity are consistent with the high white dwarf mass implied by the peak temperature (see also Section~\ref{disc}). Although a hard component is definitely present, with a count rate of a few$\times 10^{-4}$ cps above 1.5 keV, its low flux meant that meaningful investigation of the evolution of its flux or temperature was rendered impossible. Parameterising the hard spectral component as optically thin emission at an assumed temperature of 5 keV, the 90\% upper limit
on the unabsorbed 0.3-10 keV X-ray luminosity is $\sim3-8\times10^{34}$ erg s$^{-1}$ between days 74 and 170. 

\subsection{Periodic oscillations}\label{oscillations}

\begin{figure}
\includegraphics[angle=270,trim=70 10 40 80,clip,width=\columnwidth]{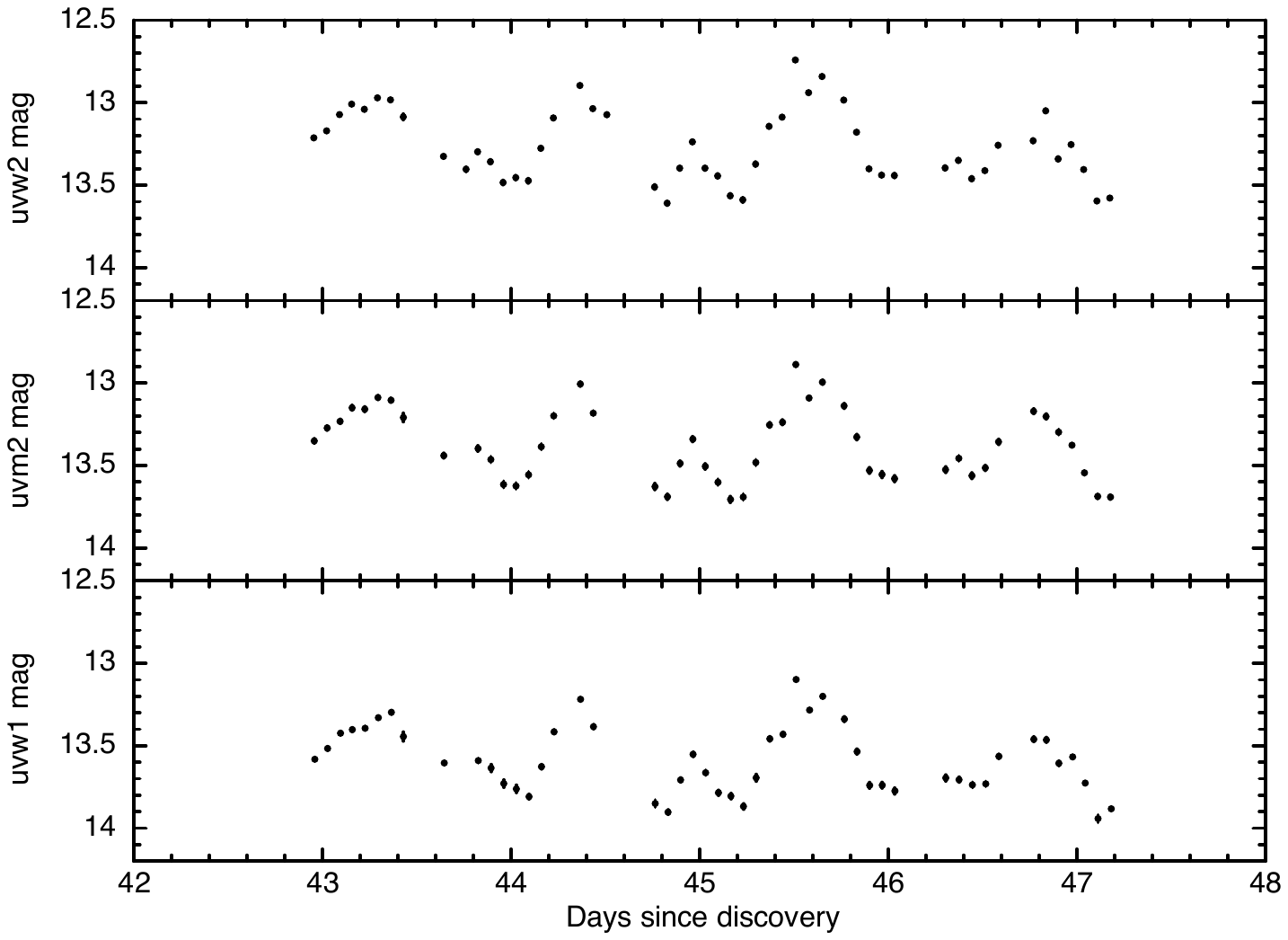}
\caption{UV magnitudes from the first set of high cadence {\it Swift} UVOT observations. Four intensity peaks are separated by $\sim 1.2$ days.\label{fig14}}
\end{figure}

As noted in Section~\ref{swift}, periodic oscillations were evident in the UVOT data from day 43. To explore this further, several sets of high cadence data were secured. Figure~\ref{fig14} shows the UVOT photometry between days 43 to 47, where four peaks separated by $\sim1.2$ days are evident.

\begin{figure}
\includegraphics[angle=270,trim=70 10 35 80,clip,width=\columnwidth]{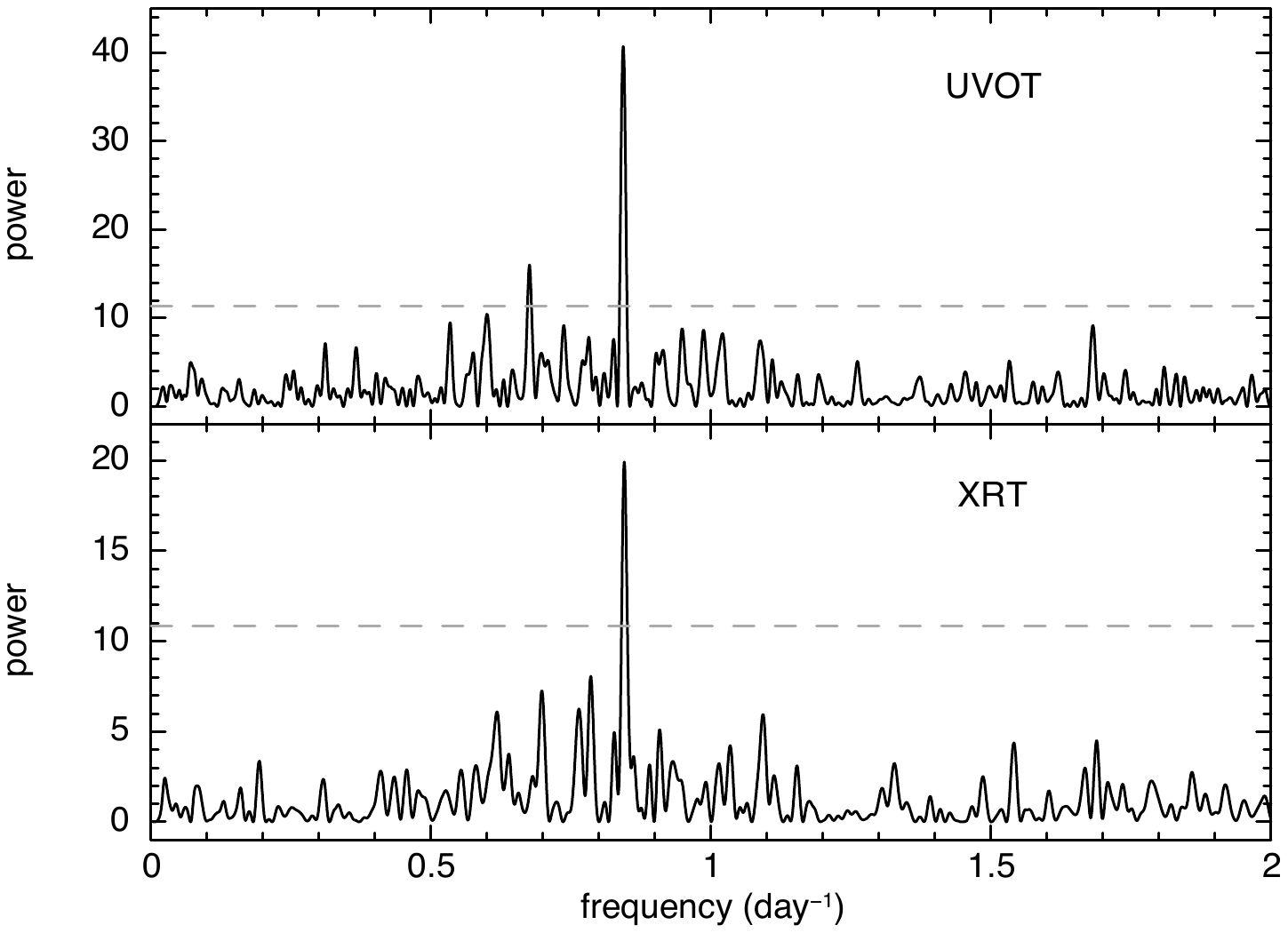}
\caption{Lomb-Scargle periodograms for the UVOT and XRT data collected between days 80 and 200. The dashed horizontal line in each panel shows the $3\sigma$ detection level \citep[i.e. the 1\% false alarm probability level due to noise, e.g.][]{sca82, hor86}. A strong modulation is detected in the UVOT data at a period of $1.1897 \pm 0.0042$ day (upper panel) and similarly in the XRT data with a period of $1.1832 \pm 0.0078$ day.\label{fig15}}
\end{figure}

The period was derived from a Lomb-Scargle periodogram of the Swift-UVOT light curve, using data out to day 200, after first detrending it by subtracting a 4th order polynomial least-squares fit (Fig.~\ref{fig15}). At early times, data were collected using all three UVOT UV filters; these were normalized to the uvw2 filter prior to periodogram analysis.

The time of phase zero was obtained by fitting a sinusoidal profile to the maximum seen in the folded UVOT light curve, excluding 0.3 cycles around the minimum where a secondary maximum is sometimes evident. The resulting ephemeris, giving the time of UV flux maximum, is

\begin{equation}\label{ephemeris}
t_{\rm max} = {\rm HJD} 2454913.174(15) + 1.1897(42) {\rm N}
 \end{equation}

where the uncertainties are $1\sigma$ estimates \citep[see for example,][]{lar96}. A Lomb-Scargle periodogram of the XRT data is shown in the lower panel of Fig~\ref{fig15}, clearly demonstrating that a modulation is also detected in the X-rays, at a period of $1.1832 \pm 0.0078$ day (using data between day 80--200).

\begin{figure*}
\begin{minipage}[t]{0.49\textwidth}
\includegraphics[trim=10 40 70 80,clip,width=\textwidth]{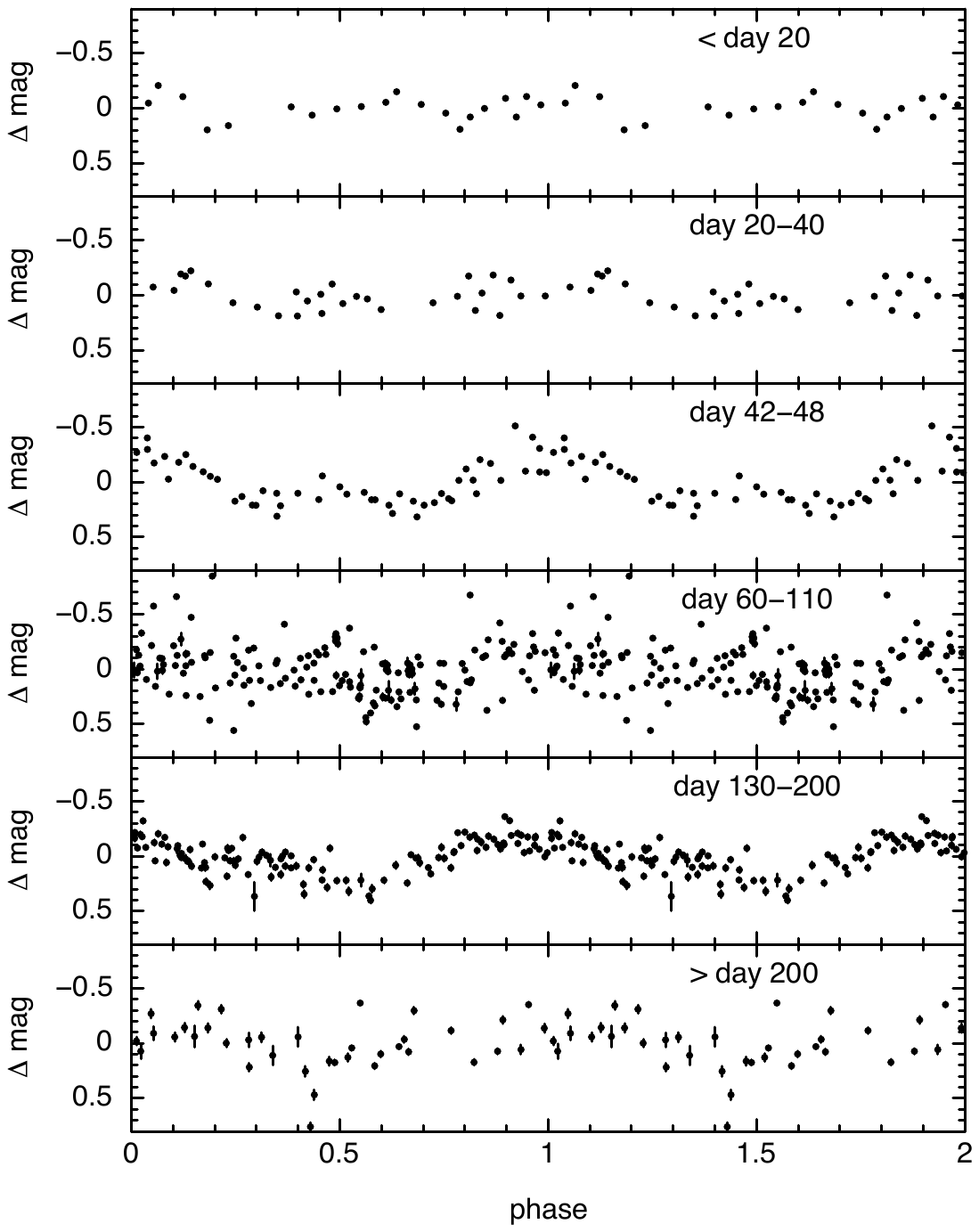}
\end{minipage}\hfill
\begin{minipage}[t]{0.49\textwidth}
\includegraphics[trim=10 40 70 80,clip,width=\textwidth]{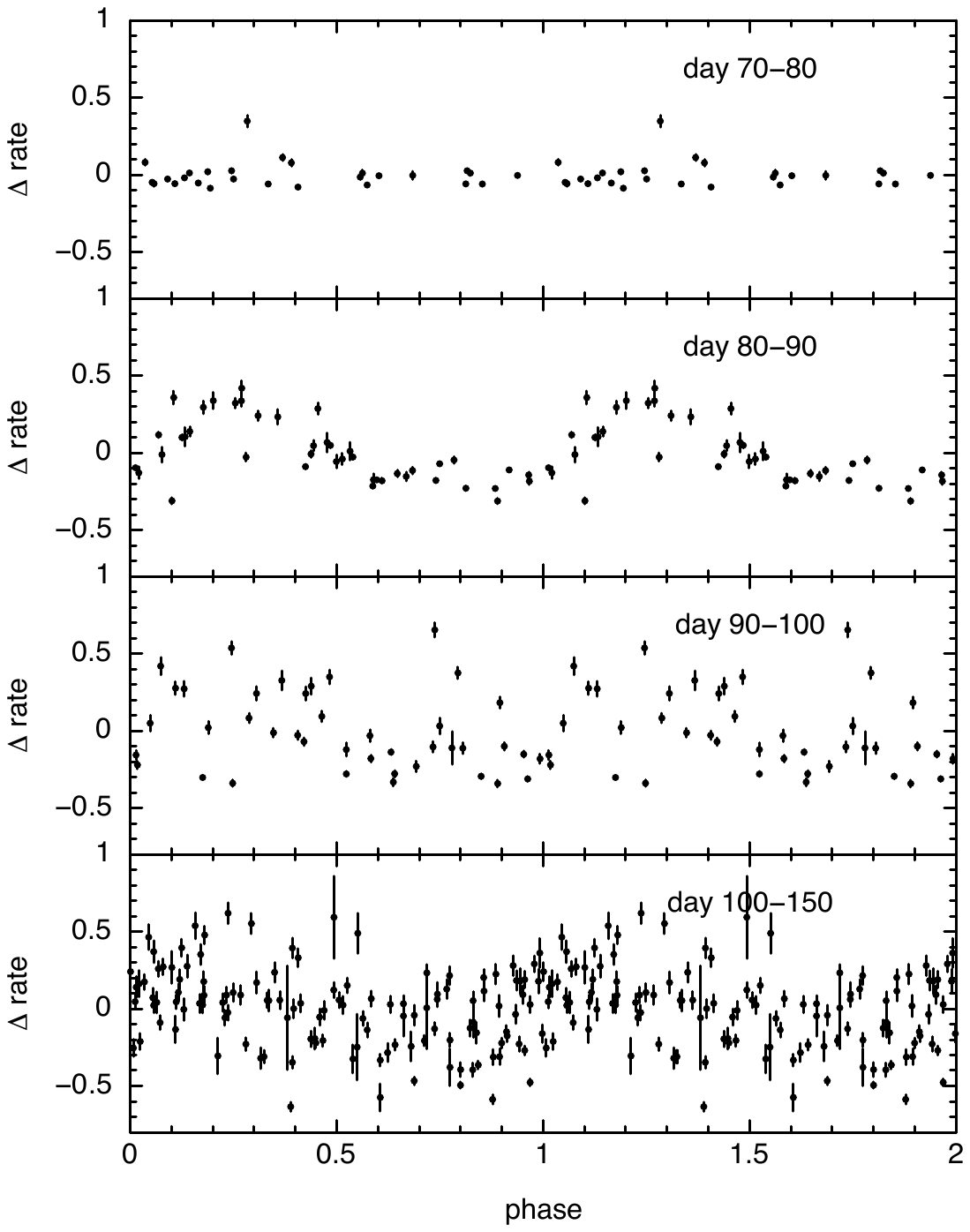}
\end{minipage}
\caption{Time-sliced UVOT (left) and XRT data (right), de-trended using a 4th order polynomial and phase-folded on the ephemeris given in Equation \ref{ephemeris} in the text. It can be seen that the modulation strength varies over time, and there is a clear difference in phase between the UV and X-ray data.\label{fig16}}
\end{figure*}

Figure~\ref{fig16} shows the UVOT and XRT data phase-folded as a function of time. For the UV data, the modulation is clearly visible between day 42--48 and 130--200, while the X-rays are most coherently modulated between day 80--90. The X-ray modulated flux peaks $0.28$ cycles (corresponding to $\sim0.3$ day) later than the UV. 

\begin{figure}
\includegraphics[angle=270,trim=70 10 40 80,width=\columnwidth]{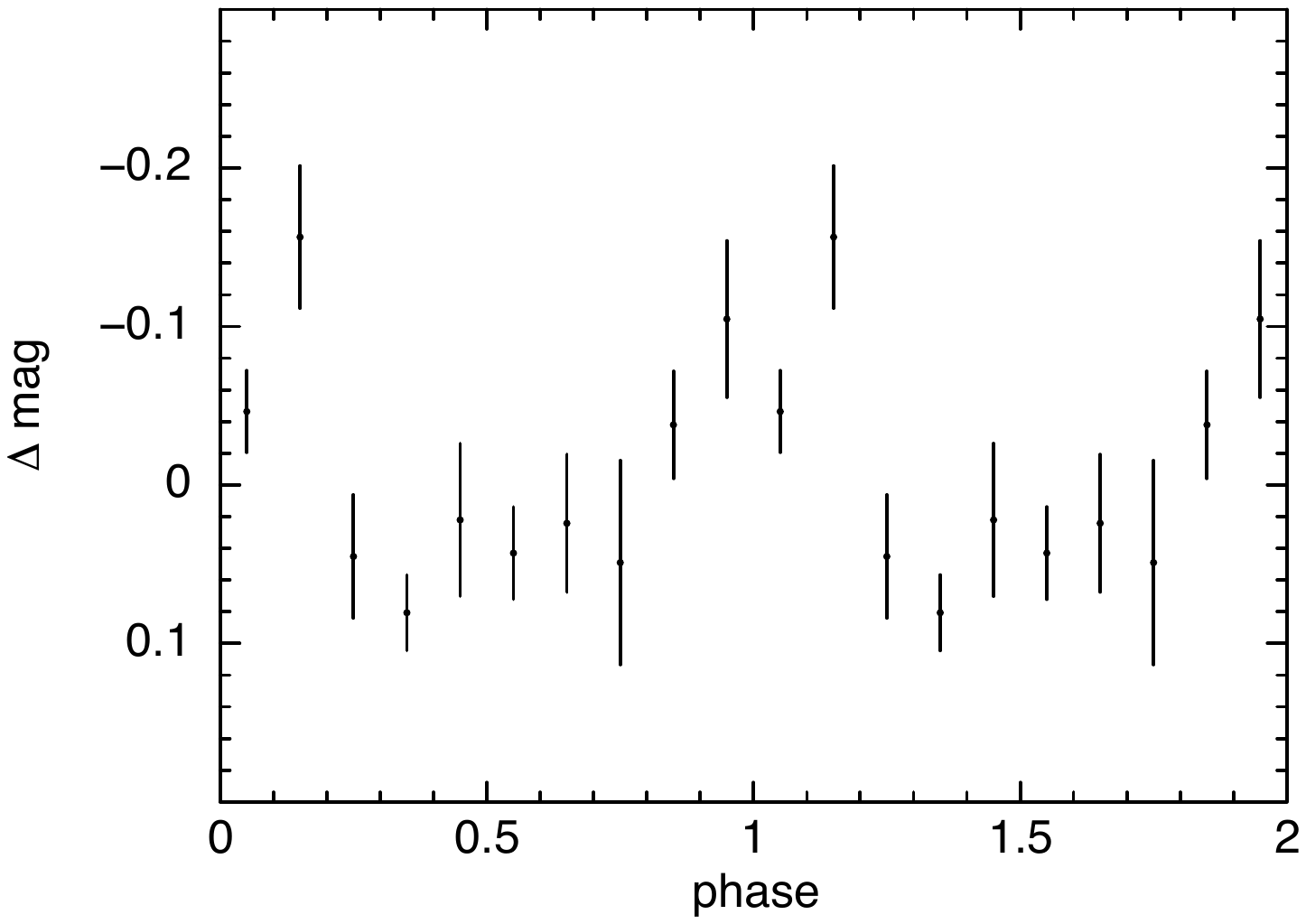}
\caption{The SMARTS $V$~band data from the first 120 days following discovery, folded on the 1.1897 day period found in the UVOT data, detrended with a 4th order polynomial. Data have been binned; the uncertainty in a bin is the standard deviation of the points in that bin. \label{fig17}}
\end{figure}

We also examined the SMARTS data for an optical counterpart to the modulation seen in the UV light curve. We recovered a sinusoidal modulation in the $V$-band data over the first 120 days (Fig.~\ref{fig17}). The phasing is that seen in the UV, within the uncertainties.

\subsection{Progenitor System}\label{quiescence}

Based on the SMARTS photometry of the 2009 eruption, we assumed that the system had returned to quiescence from 2010 March (2011 April for the NIR data), see Fig.~\ref{fig2} and Table~\ref{progen_phot}.  We also undertook a search of archival datasets to attempt to recover the system before the 2009 eruption.

\begin{figure}
\includegraphics[width=\columnwidth]{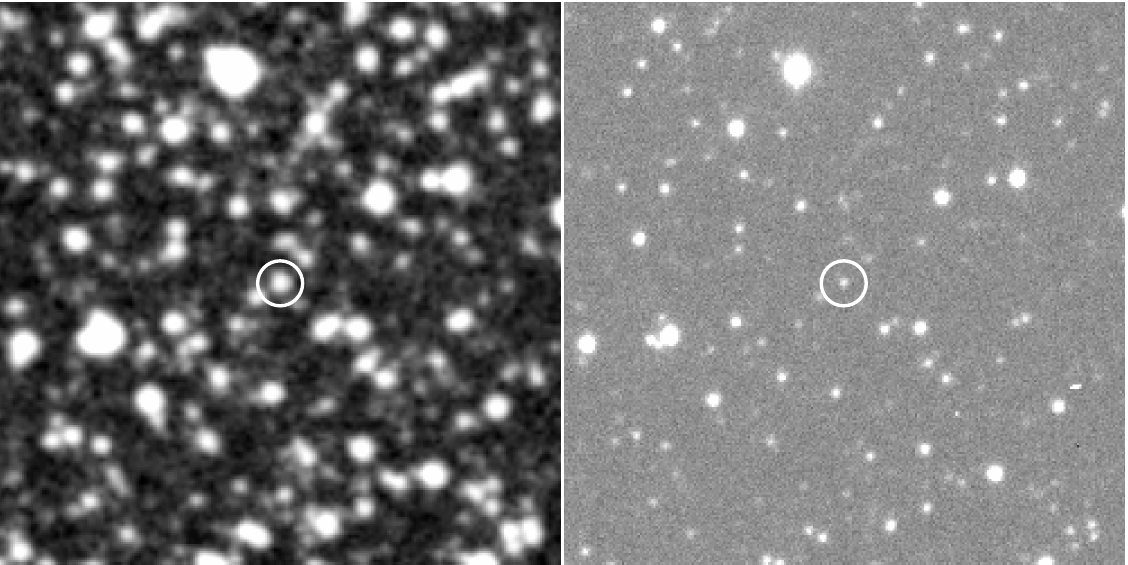}
\caption{The progenitor of nova LMC 2009a found on an SERC\_J\_DSS1 image taken on 1975 January 10 UT (left) compared with a SMARTS V-band image taken on 2009 December 28 UT (day 326 after discovery; right).\label{fig18}}
\end{figure}

We computed the spatial transformation between a $V$-band SMARTS image of the nova in eruption, taken on 2009 December 28 (day 326) and the same field from the SERC-J$_{\rm DSS1}$ archival survey (plate taken 1975 January 10) using 21 stars resolved and unsaturated in both images (see Fig.~\ref{fig18}). This approach, being independent of astrometric calibration, yields the most accurate results. The uncertainty in the derived transformations (0.26 pixels, equivalent to $0^{\prime\prime}\!\!.1$) dominates over the average positional error of the nova in the SMARTS image.  There is a resolved object within 0.14 pixels (equivalent to $0^{\prime\prime}\!\!.05$, $0.4\sigma$) seen in the archival data.  Given the distribution of objects within both datasets, the probability of finding an object at least as close to the position of the nova by chance is $<0.01$~percent \citep[following the methodology outlined in][]{2009ApJ...705.1056B}. We conclude therefore that this is the progenitor system.

Having identified the pre-eruption progenitor system, a search through other archival images was made. Whilst present in the SERC-J data, the object is out of the field of view for the SERC-I and SERC-V surveys. However, it was present on the ESO-R (1986 February 03 UT) and ESO-B (1991 January 17 UT) images (in addition to SERC-J on 1975 January 16). Five stars in the field around the nova were selected and the $B$ and $R$ magnitudes of those objects were found in the USNO-B1 catalog. Differential photometry with the GAIA\footnote{\url{http://star-www.dur.ac.uk/\textasciitilde pdraper/gaia/gaia.htx/index.html}} data analysis package was performed to find the magnitudes of the quiescent nova system as $B = 18.32 \pm 0.37$, $R = 19.20 \pm 0.62$, and $B_{\mathrm{J}} = 18.17 \pm 0.20$ (see also Table~\ref{progen_phot}).

The 2MASS on-line catalog yielded no source at, or near, the position of Nova LMC 2009a, and a re-analysis of the 2MASS $JHK_{\mathrm{S}}$ data did not detect any source at the position.  Based on the photometry of nearby faint, just-resolved, sources in the 2MASS data we derived the following $3\sigma$ upper limits for the NIR photometry of the progenitor system $J>19.3$, $H>18.2$, and $K_{\mathrm{S}}>18.7$ (see Table~\ref{progen_phot}).

The 2MASS upper limits, the SERC-J photometry, and the ESO photometry are also plotted in the SED evolution plot, see Fig.~\ref{fig4}.

\begin{figure}
\includegraphics[width=\columnwidth]{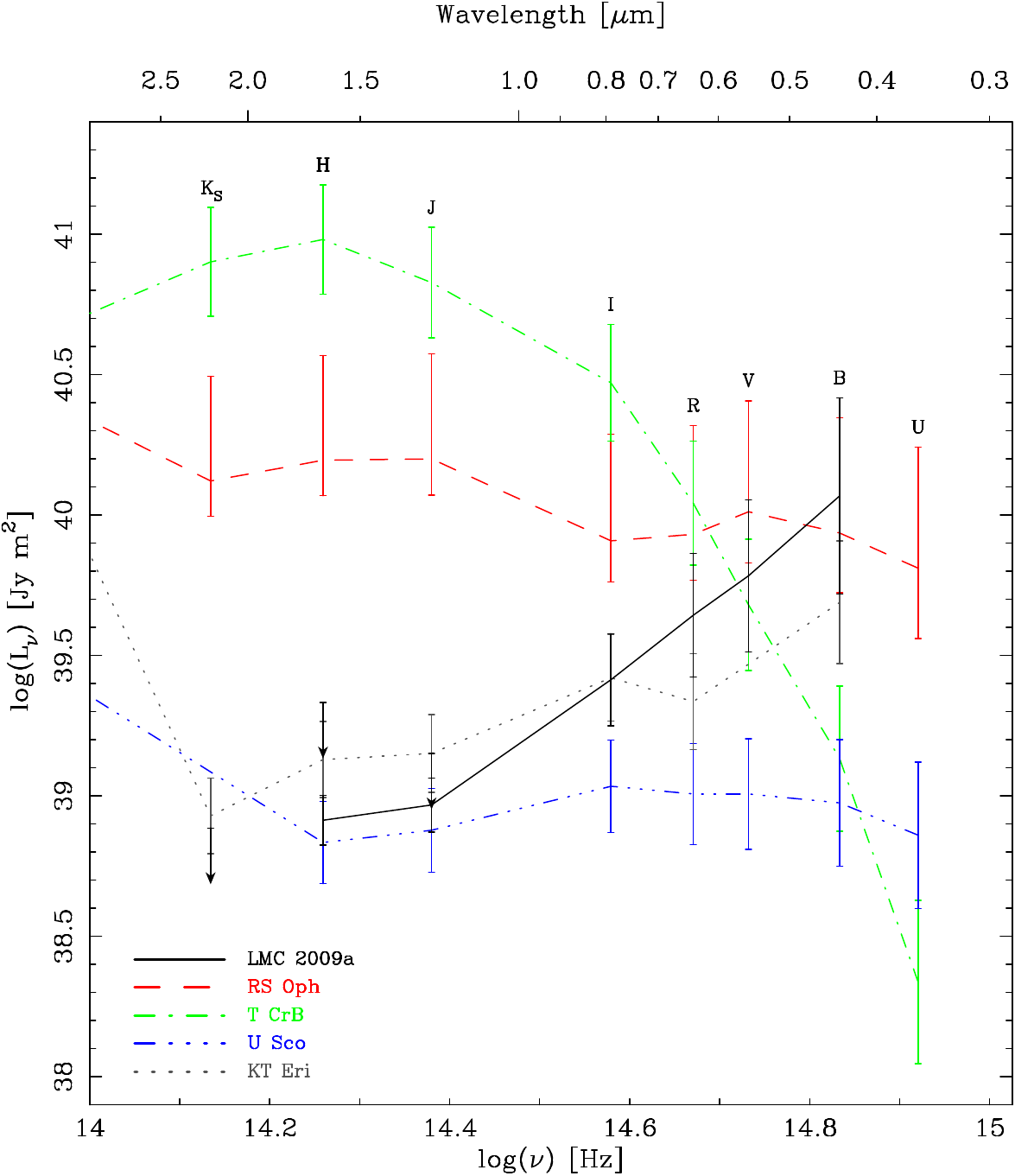}
\caption{{\scriptsize Distance and extinction-corrected SEDs for the progenitor system of Nova LMC 2009a compared to those of the quiescent Galactic RNe RS~Oph, T~CrB, and U~Sco, as well as the suspected Galactic RN KT~Eri (see Key for object identifications).  Units chosen to allow comparison with similar plots in \citet[see their Fig.~71]{schae10} and \citet[see their Fig.~4]{dar14}.  For each system, point-to-point uncertainties are relatively small, indicated error bars are dominated by systematic distance and extinction uncertainties.  Extension redwards of the $K$-band is from {\it WISE} data \citep{2014MNRAS.444.1683E}.\label{fig19}}}
\end{figure}

In Fig.~\ref{fig19} we present the distance and extinction-corrected spectral energy distribution (SED) of the quiescent Nova LMC 2009a based on the SMARTS and 2MASS photometry, those of the Galactic RNe RS~Oph and T~CrB (RG-novae), and U~Sco (SG-nova), and that of the suspected Galactic recurrent RG-nova KT~Eri.  Photometry for the Galactic RNe is derived from \citet[see their Table~30]{schae10}, and has been extended redwards of the $K$-band using {\it WISE} data \citep{2014MNRAS.444.1683E}; distances and extinction from \citet[see their Table~2 and references therein]{dar12}; additional RS~Oph photometry from \citet[and the Liverpool Telescope; \citealp{ste04}]{dar08}; KT~Eri photometry is from \citet{jor11}, and optical and NIR absolute calibrations from \citet{bes79} and \citet{cam85}, respectively.  We assume a distance to RS~Oph of $d =1.4^{+0.6}_{-0.2}$~kpc (\citealp{bar08}; see also \citealp{bod87}).

\begin{figure*}
\begin{minipage}[t]{0.49\textwidth}
\includegraphics[width=\textwidth]{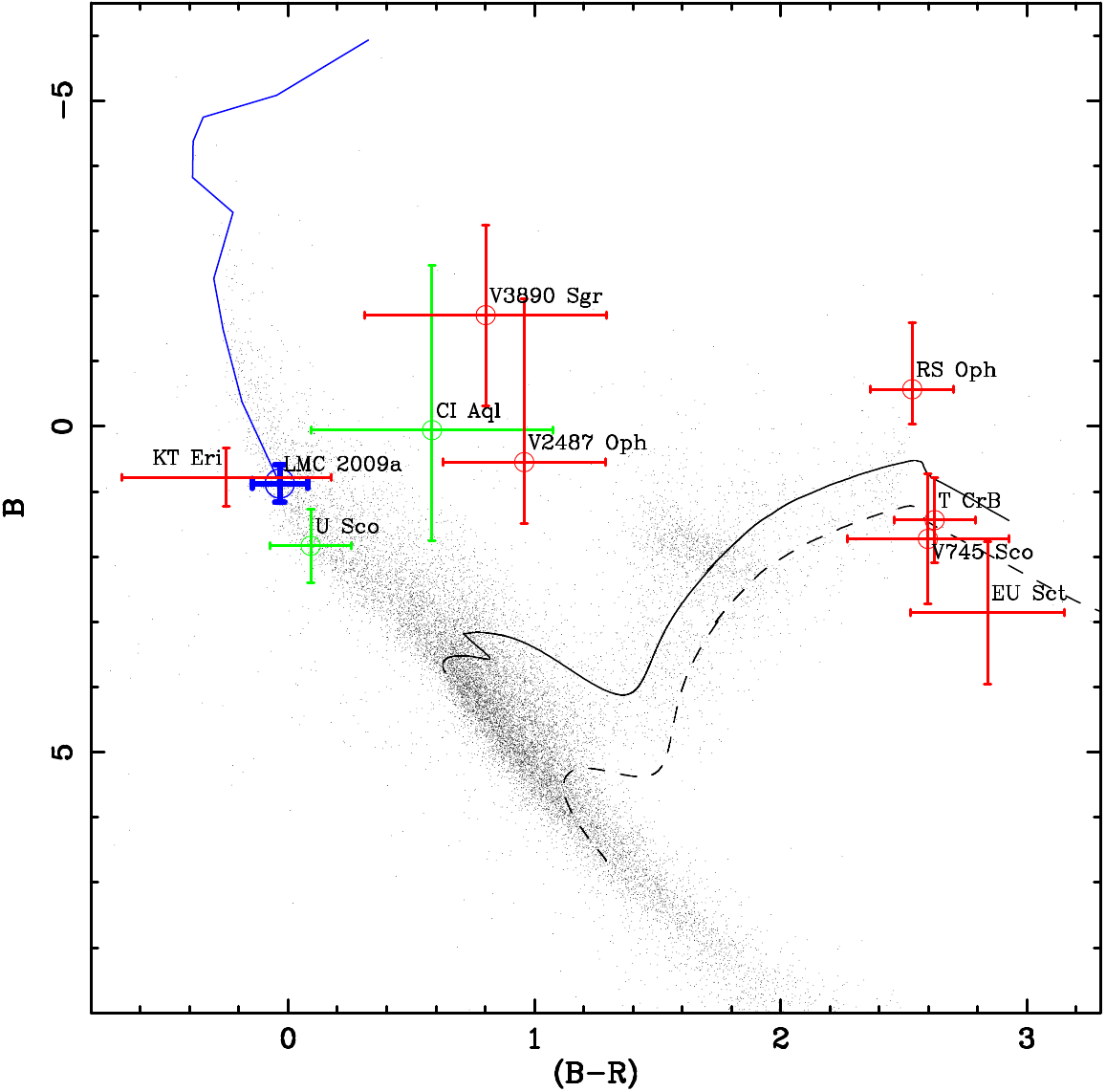}
\end{minipage}\hfill
\begin{minipage}[t]{0.49\textwidth}
\includegraphics[width=\textwidth]{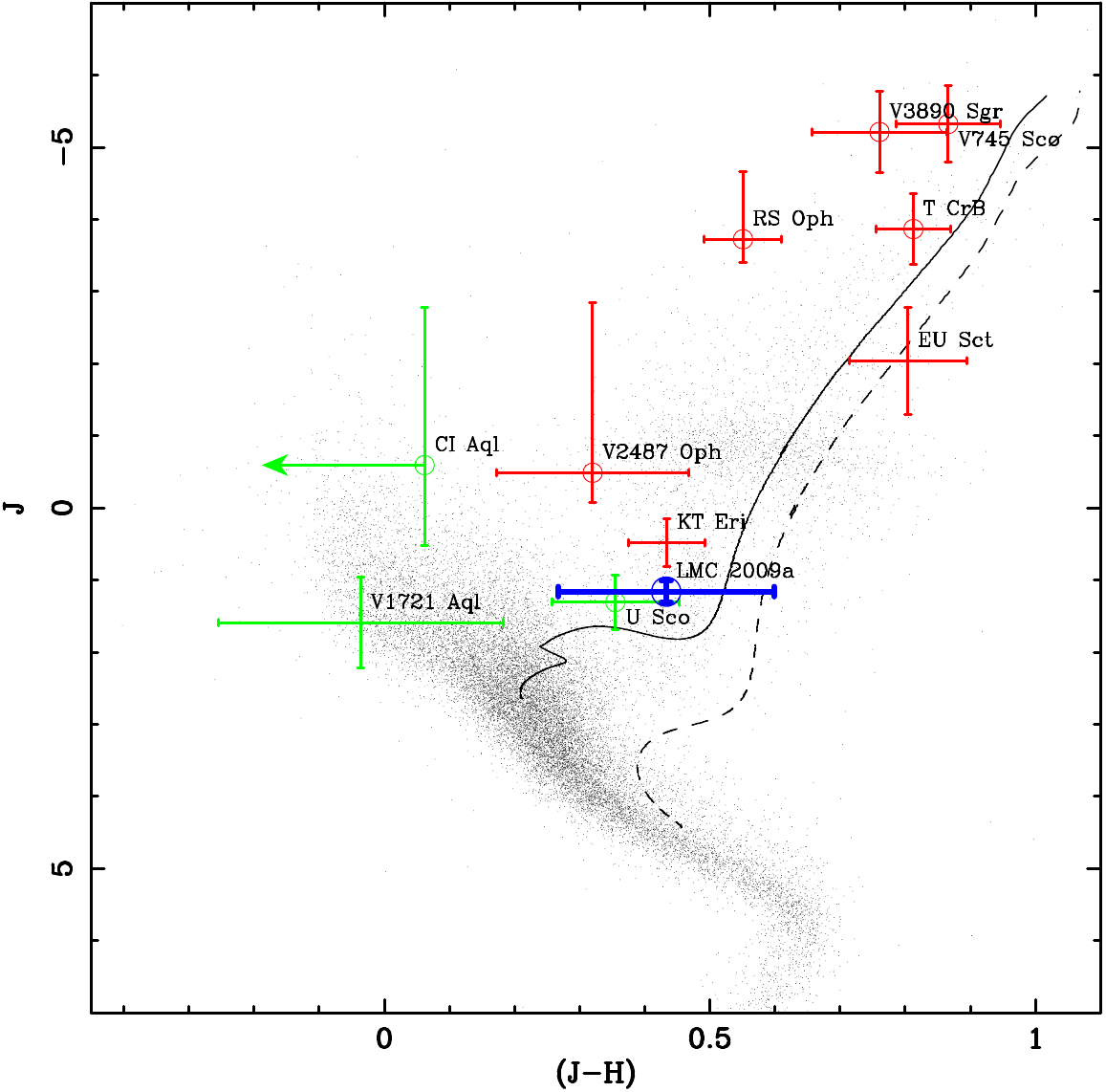}
\end{minipage}
\caption{Color-magnitude diagrams showing stars from the {\it Hipparcos} data set \citep[black points;][]{per97} with parallax errors $< 10$~percent.  These stars have been transformed to the distance and extinction of the LMC as given in the text.  $B$ photometry is taken directly from the {\it Hipparcos} catalogue, $R$ photometry is taken from the NOMAD1 data set (\citealp{zac05}; via the VizieR database, \citealt{och00}), and NIR photometry from the 2MASS catalogue.  The dark blue data point shows the location of the nova LMC~2009a quiescent system, considering the uncertainty in the photometry and extinction.  The dark blue line indicates the track of the nova from day 10 post-eruption back to quiescence.  
The hashed-black and solid-black lines are evolutionary tracks of $1 M_\odot$ and $1.4 M_\odot$ stars, respectively \citep{pie04}.  The red points represent Galactic RG-novae and the green points Galactic SG-novae \cite[see][and references therein]{schae10,dar12}.  The known recurrent novae in this sample have been identified by an additional circle. The $B$ vs.\ $(B-R)$ color-magnitude diagram is very sensitive to the the accretion disk in the system, hence the position of LMC 2009a (similar to a massive main sequence star) implies the presence of a luminous and hot disk.  The $J$ vs. $(J-H)$ diagram is much less sensitive (in general) to the disk, and more so to the secondary; here the position of LMC 2009a lies around the sub-giant branch or the base of the red giant branch.
\label{fig20}}
\end{figure*}

Figure \ref{fig20} presents a $B$ versus $(B-R)$ and a $J$ versus $(J-H)$ color-magnitude diagram (CMD) showing stars from the {\it Hipparcos} data set \citep{per97} merged with the NOMAD1 \citep{zac05} and 2MASS data sets, respectively, via the VizieR database \citep{och00}, with parallax errors $< 10$~percent.  These stars have been translated to the distance, and estimated extinction, towards the LMC. We discuss the conclusions to be drawn from the data shown in Figs \ref{fig19} and \ref{fig20} in the next section.

\section{Discussion}\label{disc}

\subsection{Optical parameters}\label{optical_parameters}

As the distance to the nova is reasonably well constrained (see Section \ref{SMARTS_photom}) compared to the typical case for Galactic novae, we can derive a fairly robust estimate of its absolute magnitude at peak. The main uncertainty lies in the extinction. Using $E(B-V) = N_{\rm H}/ (6\pm2)\times10^{21}$ cm$^{-2}$, with  $N_{\rm H} = 1.13\times10^{21}$cm$^{-2}$  \citep[see Section~\ref{swift} and][]{kal05} and $A_V = 3.1 E(B-V)$, $A_V = 0.6\pm0.2$. The online LMC extinction calculator\footnote{\url{http://djuma.as.arizona.edu/\textasciitilde dennis/lmcext.html}} derived from \cite{zar04} gives $<A_V>$ = 0.31$\pm0.30$ from the results for 22 stars within $1'$ of the position of the nova. This is at least consistent with the result from the H column. Almost all of this extinction lies within the LMC itself \citep[][note that the foreground Galactic extinction is minimal - $A_V \sim 0.05$]{zar04}.

The peak magnitude of the eruption is subject to at least two uncertainties. First of all, there is only a single observed magnitude on the night of discovery (SMARTS observations beginning the next night) and this is an unfiltered magnitude, compared to those obtained subsequently by us. However, if we take $[m_V]_{\rm max} = 10.6\pm0.1$ and the values of $d$ and $A_V$ above, $M_V = -8.4\pm0.8_r\pm0.7_s$. As an aside, we note that this is very similar to $M_V = -8.7$ derived for the Galactic nova KT~Eridani by \cite{rag09} which is discussed further below. Using the MMRD relationship from \cite{del95}, with $t_2 = 5$ days for LMC 2009a (see Table \ref{tbl-phdecay}), the implied $M_V = -8.9$, which is comparable to the observed peak magnitude, within the quoted errors. It should be noted that \cite{del95} use a fit to novae observed in M31 where the intrinsic scatter due to distance uncertainties is less than for Galactic novae and the sample includes several novae with declines at least as fast as that of LMC 2009a, again unlike the linear MMRD relationships derived for Galactic novae (e.g. \citealt{dow00}; we note however that \citealt{kas11} have cast some doubt on the overall validity of the MMRD relations).

Although the peak absolute magnitude is consistent with those found for CNe of a similar speed class, Nova LMC 2009a would appear anomalous in terms of the amplitude of the eruption, $A$. We estimate $A \sim 9$ magnitudes in the $V$-band. This compares to $A \gtrsim 13$ for a  CN of this speed class from the $A$ vs $t_2$ relationship for CNe given in \cite{war08}. Such a low amplitude of eruption implies either (i) the peak luminosity is much lower than that of typical of CNe or (ii) the luminosity of the  quiescent system is much higher than is typical in such objects. As the peak absolute magnitude does not appear anomalous, either the secondary star is more luminous than is normal in a CN at quiescence, or there is a greater contribution from the accretion disk (assuming emission from the WD itself is negligible in the optical - see further discussion below), or a combination of the two. Such low amplitude eruptions are however typical of RNe due to their relatively high quiescent luminosity.

\subsection{Parameters of the central system}\label{parameters_central_system}

\subsubsection{SSS duration and WD mass}\label{SSS_duration}

It is well established that the SSS arises from continued nuclear burning on the WD surface following the TNR which is gradually unveiled as the mass loss rate declines and the ejecta move outwards, leading to a decrease in optical depth in X-rays \citep{kra08}. The deduced temperature and luminosity of the SSS in the case of LMC 2009a are consistent with this model (see Fig.~\ref{fig13}). Simplistically, the timescale for uncovering and observed onset of the SSS phase is given by $t_{\mathrm{on}} \propto M_{\rm H}^{1/2} v_{ej}^{-1}$ \citep{kra96} where $M_{\rm H}$ is the mass of H in the ejected envelope and $v_{\mathrm{ej}}$ is the ejection velocity. Thus for the low ejected masses and high ejecta velocities found in  RS~Oph-type and U~Sco-type RNe, $t_{\mathrm{on}}$ would be expected to be relatively short compared to most CNe.  

The turn-off time since eruption for nuclear burning, $t_{\rm rem}$, is a steep function of WD mass;  \cite{mac96} finds for example  $t_{\rm rem} \propto M_{\rm WD}^{-6.3}$. Similarly, the timescale after eruption for the onset of the SSS phase, $t_{\rm on}$, is also a function of $M_{\rm WD}$ in the sense that $t_{\rm on}$ is likely to be shorter for systems containing a high mass WD. As noted above, both U Sco and RS Oph have a very short observed $t_{\rm on}$ of $\lesssim 12$ days \citep{sch10} and $\lesssim 30$ days \citep{bod06,osb11} respectively. In both cases, the WD mass has been concluded to be approaching the Chandrasekhar mass limit of M$_{\rm Ch} \sim 1.4$M$_{\odot}$ \citep{kah99,hac07,osb11}. When considering the rate of SSS evolution, the most extreme example of a recurrent nova is V745 Sco. This nova entered the SSS phase only four days after eruption, with $t_{\rm rem}$ as short as 6 days \citep{pag14,pag15}. Here the WD mass has not been estimated in detail as yet, but is again likely to be high. In terms of extragalactic novae (and in terms of recurrence timescale for any RN), the most extreme example of an RN is M31N~2008-12a with eruptions suspected to occur at least annually \citep{hen15b} and extremely rapid SSS evolution \citep[][$t_{\rm on} < 6$ days and $t_{\rm rem}  \sim 18$ days]{dar14, dar15, hen14, hen15, kat14, kat15, tan14}. Again, the WD mass is concluded to be very near M$_{\rm Ch}$ and here the accretion rate is found as $\dot{M}_{\rm acc} > 10^{-7} M_\odot$ yr$^{-1}$, which is extremely high for CVs, including old novae  \citep[e.g.][]{pat84}.

For LMC 2009a, $t_{\rm on} \lesssim 70$ days \citep{bod09b} and $t_{\rm rem} \simeq 270 \pm 10$ days (from Fig.~\ref{fig11}). The fact that $t_{\rm on}$ is much later in LMC 2009a is consistent with the lower observed ejecta velocities in this object compared to either U Sco or RS Oph, and may also indicate a higher ejected mass. In the case of $t_{\rm rem} = 270$ days, relationships in \cite{sta91}  imply $M_{\rm WD} = 1.3 M_\odot$ in LMC 2009a. Similarly, from the relationship quoted above from \cite{mac96}, using $t_{\rm rem} = 58$ days and $M_{\rm WD} = 1.35 M_\odot$ in RS Oph \citep{osb11}, $M_{\rm WD} = 1.1 M_\odot$ in LMC 2009a. In addition, we note from the work of \cite{wol13} that the peak effective temperature of the SSS that is found in LMC 2009a, $T_{\rm peak} = 87\pm1$~eV, is consistent with $1.2 M_\odot \lesssim M_{\rm WD} \lesssim 1.3 M_\odot$ (their Fig. 10). Thus we conclude that here $1.1 M_\odot \lesssim M_{\rm WD} \lesssim 1.3 M_\odot$.

\subsubsection{Mean mass accretion rate}\label{accretion_rate}

The results of a grid of nova models by \cite{yar05} illustrate the fact that for an interval between eruptions $\Delta t_{\rm eruption} \sim 38$ yr, both $M_{\rm WD}$ and the mean mass accretion rate $\dot{M}_{\rm acc}$ need to be high (we note this as the mean rate as e.g. \citealt{shav14} show that $\dot{M}_{\rm acc}$ may be a highly variable parameter). We may estimate the mass of the accreted envelope, $M_{\rm acc} $, required to give rise to a TNR from 

\begin{equation}\label{accretion}
M_{\rm acc} \simeq \frac{4\pi R_{\rm WD}^{4}}{{\rm G} M_{\rm WD}} P_{\rm crit},
\end{equation}

\noindent\citep[see e.g.][]{sha81,sta89}

If we take $M_{\rm WD} = 1.2 M_\odot$ then $R_{\rm WD} = 4.3 \times 10^8$ cm \citep[from fitting to the formula for the Chandrasekhar mass-radius relation by Eggleton, and reported in][]{tru86}. We note that this is of the same order as the radius of the emitting region derived from the model atmosphere fits - see Fig.~\ref{fig13}. With $P_{\rm crit} \simeq 10^{19}$ dyne cm$^{-2}$ \citep{yar05}, then $M_{\rm acc} \simeq 1.4 \times 10^{-5} M_\odot$. Thus with $\Delta t_{\rm eruption} \sim 38$ years, $\dot{\it{M}}_{\rm acc} \simeq 3.6^{+4.7}_{-2.5} \times 10^{-7} M_\odot$ yr$^{-1}$ for the deduced range of WD mass, which as noted above is extremely high (we may also note that \cite{tru86} give $P_{\rm crit} \simeq 10^{20}$ dyne cm$^{-2}$ which obviously would increase the derived mass accretion rate further still, as would a lower value of $\Delta t_{\rm eruption}$ if there were any eruptions missed between 1971 and 2009 as might be the case for such a fast declining nova). 

Consulting the \cite{yar05} grid of nova eruption models, $M_{\rm WD} = 1.25 M_\odot$ with $\dot{M}_{\rm acc} = 10^{-7} M_\odot$ yr$^{-1}$ gives $\Delta t_{\rm eruption} \sim 20$ yr, which is compatible with that for LMC 2009a, although the ejection velocities they derive are much lower than those we have observed in LMC 2009a (underprediction of the ejecta velocities by current theoretical models is however well known).

\subsubsection{Origin of the periodic oscillations}\label{periodic_oscillations}

Considering the 1.2 day periodic oscillations seen in
both the UV/optical and X-rays, what is
striking is the smooth modulation of the X-ray emission
which does not go to zero flux. The period is much too
long to be due to WD rotation, so the smooth X-ray
modulation is not due to a hot spot on the WD (e.g. due
to a strong magnetic field). Orbital modulation must therefore be considered. However, eclipse of a small hot white dwarf would result in sharp flux transitions. These are not seen, so there is no eclipse
of the WD either by the secondary star in the system or likely by a smooth accretion disk bulge; a larger optically thin X-ray scattering region around the white dwarf may be implicated by the broad modulation.
Henceforth, we will assume that the observed 1.2 day modulation represents the orbital period, placing this system in the
U Sco class of RNe, and the general class of SG-novae.

The different phasing between the UV/optical and X-rays
clearly indicates a different origin of the UV/optical and
X-ray emission, it is also notable that when
there is X-ray modulation there is no UV modulation and
vice versa (on occasion). It seems likely that the UV flux is due to the reprocessing of illuminating X-rays within the binary system, but evidently the reprocessing site has a different view of the illuminating white dwarf than we do. We can at least hypothesise that an intervening disk rim may be variable in height due to changes in the nature of the flow from the secondary or due to changes in its effective illumination; this may be the cause of some of the modulation changes we have seen.

\subsubsection{Other parameters of the quiescent system and long-term behaviour}\label{other_parameters}

As can be seen in Fig.~\ref{fig2} and Fig.~\ref{fig4}, the nova returned to a long lived ($>1,800$ days) reduced luminosity level around 400 days after the eruption, which, for now, we will assume is the quiescent state.  As the optical and NUV parts of the SEDs in both Fig.~\ref{fig4} and Fig.~\ref{fig19} both indicate, the high luminosity and blue color of the emission from the system at quiescence are consistent with a system whose luminosity is most likely dominated by a bright accretion disk.  As the comparative plots in Fig.~\ref{fig19} show, the luminosity of the Nova LMC 2009a disk is significantly higher than that present in U~Sco, and on a par with that contained in the RS~Oph and KT~Eri systems.  Such a bright disk implies a high mass transfer rate, which is in turn consistent with the conclusion above from consideration of the accreted mass and inter-eruption timescale. For example, using 

\begin{equation}\label{luminosity}
L_{\rm acc} = (1 - \alpha) \frac{G M_{\rm WD} \dot{M}_{\rm acc}}{2 R_{\rm WD}},
\end{equation}

\noindent with $\alpha = 0.5$ \citep{sta88}, $M_{\rm WD} = 1.2 M_\odot$, $R_{\rm WD} = 4.3 \times 10^8$ cm and $\dot{M}_{\rm acc} = 3.6 \times 10^{-7} M_\odot$ gives $L _{\rm acc} \simeq 550 L_\odot$ (which is far less than $L_{\rm Edd}$ of course, so the accretion will not be disrupted). 

Turning to the quiescent NIR emission, the quiescent system is not detected in the 2MASS data. This and the $\sim1$ day orbital period seen in Nova LMC 2009a allow us to confidently rule out the presence of a luminous red giant secondary (akin to the RS~Oph or T~CrB systems) within the Nova LMC 2009a system (see Fig.~\ref{fig19}).

The quiescent SED shows remarkable similarity to that of the galactic RN candidate KT~Eri.  \citet{jor11} proposed that the KT~Eri system has a 737~day orbital period and contains a low luminosity (i.e.\ young) red giant secondary.  As noted above, the 1.2d period seen in the Nova LMC 2009a system implies that the secondary here is actually a sub-giant, akin to the secondary in the U~Sco system.  The quiescent NIR photometry from SMARTS is consistent with emission from a secondary of similar luminosity to that seen in the U~Sco system.

We also note that Nova LMC 2009a lies in a different region of the CMD (see Fig.~\ref{fig20}) from RS~Oph and all other RG-nova with the exception of KT~Eri, and is grouped with those systems whose optical emission is accretion disk dominated.  The location of the quiescent Nova LMC 2009a on the NIR CMD is consistent with the U~Sco system and therefore again with the secondary star being a sub-giant.

From the data presented in Fig.\ref{fig19}, the accretion disk is
in fact among the most luminous seen in any nova. This
implies an extremely high mass accretion rate, and/or a
disk close to face-on; however a very low system inclination is ruled out by our detection of the 1.2 day flux modulation, which we assume is orbital in origin. In comparing LMC 2009a with other novae in Fig.\ref{fig19}, it should
also be noted that U Sco is a high inclination, eclipsing
source \citep[and references therein]{scha11}. The
implied higher accretion luminosity in LMC 2009a compared to U Sco is also consistent with its slightly lower
observed ejection velocities, in line with the results of the
\cite{yar05} models as discussed above.

When comparing the post-2009 eruption quiescent data to the pre-2009 eruption data we first note that the ESO-R data from 1986 February (see Section~\ref{quiescence}) are consistent with the SMARTS post-eruption data.  However, both the SERC-J (1975) and ESO-B (1991) data show significantly brighter emission ($\sim2$ magnitudes brighter).  The SERC-J and ESO-B observations are both consistent with the luminosity seen $\sim 300$ days after the 2009 eruption, that is $\sim 100$ days before reaching quiescence.  There are two possibilities that could explain these apparent discrepancies.  1) it is possible that the recurrence time of this system is much shorter than the observed 38 years and that eruptions in $\sim1974$ and $\sim1990$ were missed.  However, it seems quite unlikely that two, effectively random, observations would sample essentially the same portion of the decline phase.  2) alternatively, if the accretion disk in the system is significantly disrupted by an eruption, it could take some time to return to its pre-eruption luminosity, therefore it is possible that even $2,250$ days post-eruption the disk has yet to fully `recover'.  However, the SERC-J observation took place only $\sim1,000$ days after the 1971 eruption.  The SMARTS data over nearly 2,000 days at quiescence do not imply a slow re-brightening with time, nor do they show any high amplitude flickering that could explain the archival data.  Neither of these scenarios seems to satisfactorily explain this discrepancy.  Indeed, if we take the combination of the SERC and ESO data as the true quiescent level, then the accretion disk is even more luminous than is implied by the SMARTS data.

\subsection{Comparison with other novae at outburst}\label{comparison}

\begin{figure*}
\includegraphics[angle=270, width=\textwidth]{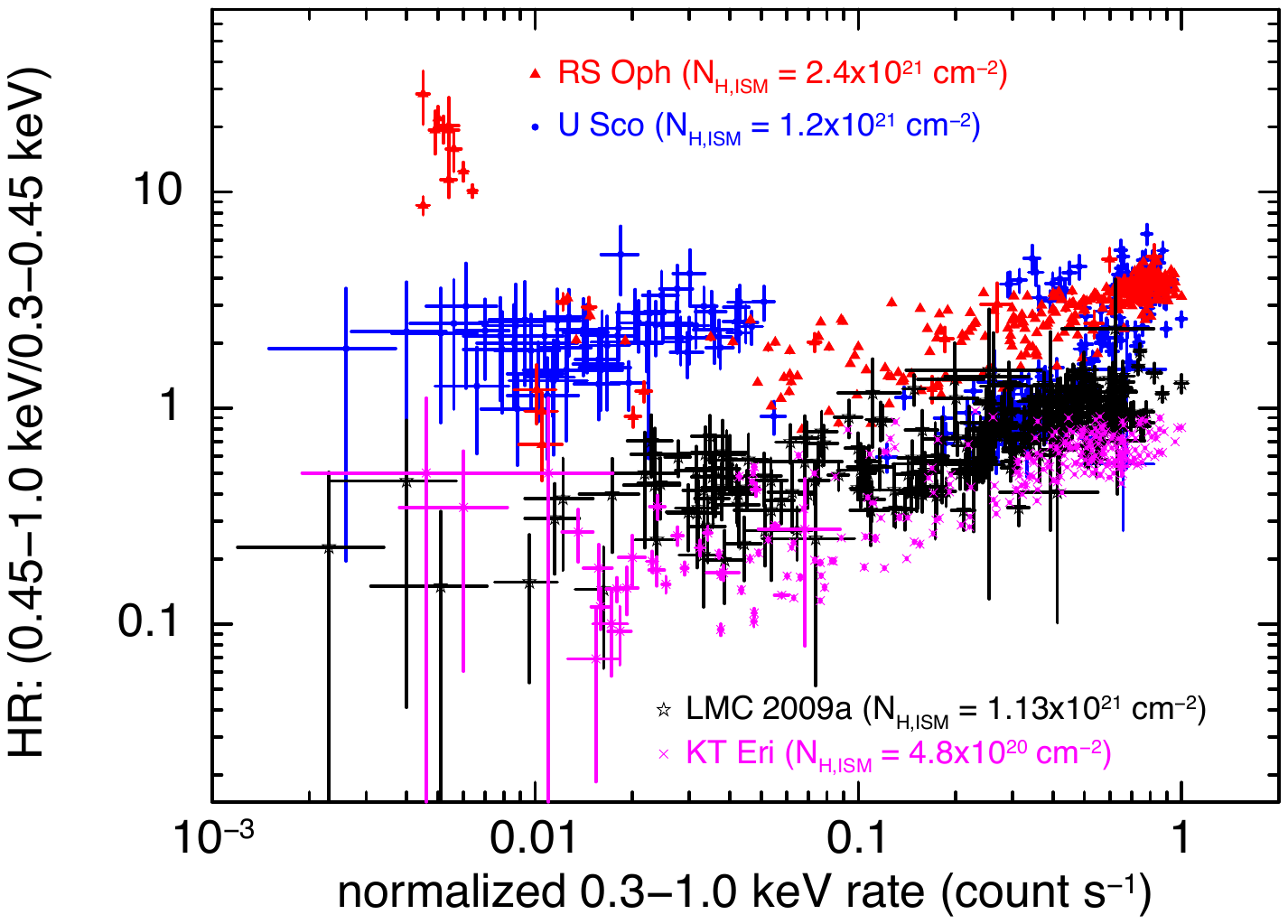}
\caption{Evolution of the X-ray hardness ratios vs XRT counts for 4 well-observed novae through the SSS phase. Note that in each case the count rate has been normalised to observed peak counts.\label{fig21}}
\end{figure*}

Fig.~\ref{fig21} shows a plot of the X-ray hardness ratio versus XRT count rate (normalised to the peak counts) for each of the novae RS~Oph, U~Sco, KT~Eri, and LMC 2009a observed in detail by {\it Swift} and showing SSS phases \citep[see also][]{sch11}.  There are several points to note stemming from this diagram.

First of all, the locus of the super-soft source phase above a value of normalised counts of $\sim 0.02$ cps is very similar for RS Oph, KT Eri and LMC 2009a (that for U Sco shows more scatter). The vertical offsets in the loci for these three novae are consistent with the differences in interstellar column noted on the figure. At low source counts, we immediately see the group of points in RS Oph at relatively high HR. These reflect emission dominated by that from the shocked wind at early and late times. The relative absence of this feature in the other novae is consistent with there being a far less dense pre-existing circumstellar envelope in these sources, in turn consistent with the presence of a less evolved secondary than in RS Oph. Similarities between the locus of points for LMC 2009a and KT Eri are also evident, and we now consider the comparison of these two objects in more detail.

\begin{figure}
\includegraphics[angle=270,trim=70 10 40 80,width=\columnwidth]{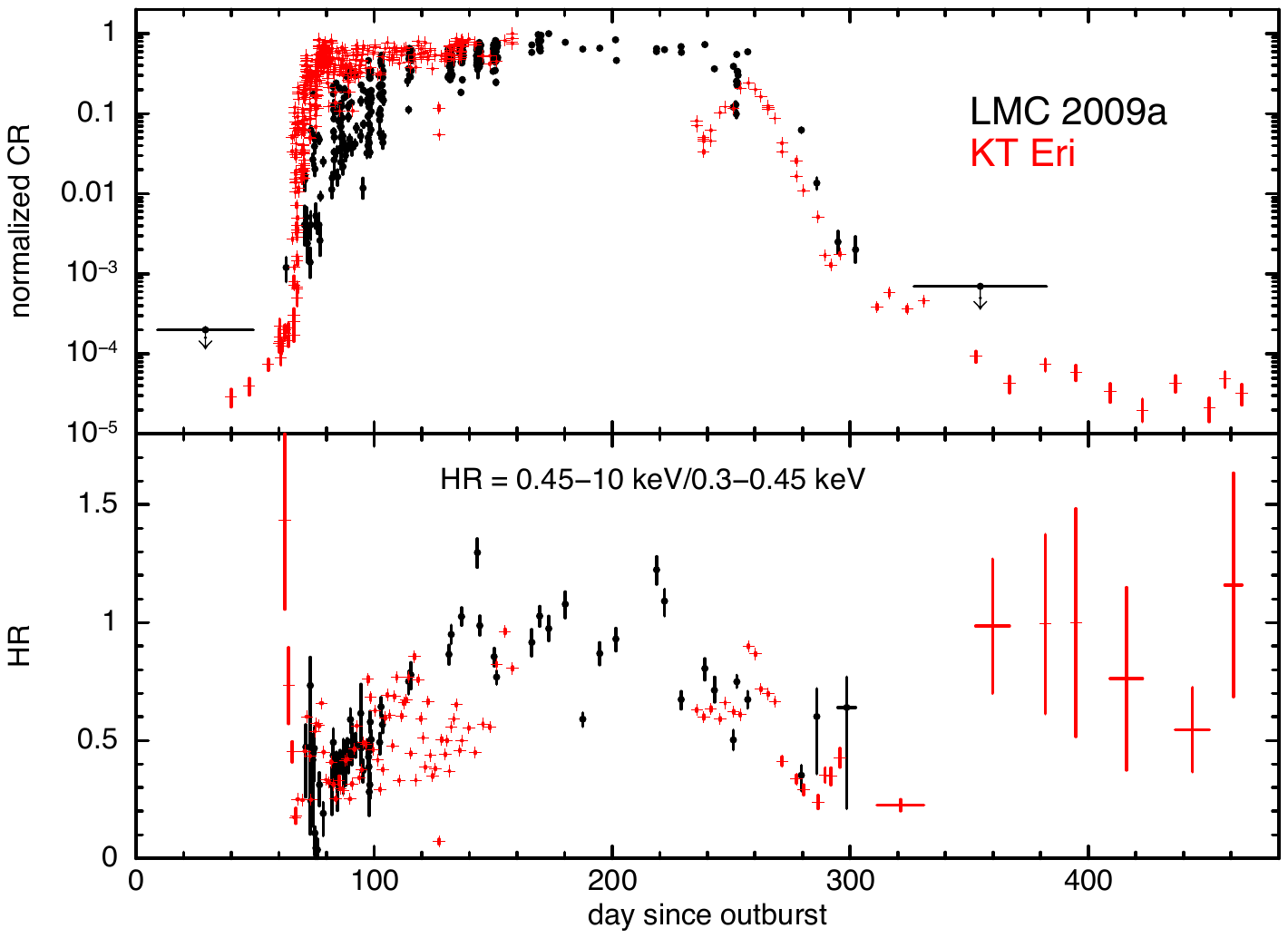}
\caption{Comparison of the X-ray evolution of Nova LMC 2009a with KT Eri (see text for details).\label{fig22}}
\end{figure}

Fig.~\ref{fig22} shows the evolution of the observed XRT counts (normalised to peak counts) and HR for LMC 2009a and KT~Eri. It is immediately apparent that the behavior of LMC 2009a is very similar to that of KT Eri on this plot. Indeed, the evolution of the optical spectra in these two novae is also very similar (Walter et al. in preparation).   Although the first detection of the SSS in KT Eri was at 55.5 days post-eruption \citep{bode10}, this would not have been detected at the distance of LMC 2009a and the dramatic rise in counts in KT Eri was observed on day 65.7, which compares to the detection of the SSS in LMC 2009a at $t \lesssim 70$ days \citep{bod09b}.  Similarly, although the later time XRT light curve of KT Eri shows some complexity, the start of the final decline of the SSS appears to be at $\sim 260$ days, compared to $270 \pm 10$ days for LMC 2009a as given above. Furthermore, taking into account the different distances of KT Eri \citep[6.5 kpc,][]{rag09} and LMC 2009a (48.1 kpc), the peak SSS count rates are within a factor of 2 of one another, which may be accounted for at least in part by the lower column to KT Eri \citep[$N_H \simeq 5 \times 10^{20}$ cm$^{-2}$;][see also Fig \ref{fig20}]{rag09,bode10}. Finally, the amplitudes of the eruptions in KT Eri and LMC 2009a $(A \sim 8-9$ mag in $V$) are also very similar. There are some similarities in the quiescent SEDs of the two novae as noted above, and intriguingly short period quasi-periodic oscillations have been detected in both novae with $P = 33$s in LMC 2009a \citep[with XMM;][]{nes14, nes15} and 35s in KT Eri \citep[with Swift;][]{bea10, nes15}\footnote{We note that scaling relations between light curve time scales (and related parameters) of different novae were also found in theoretical models \citep[see e.g.][]{hac10} and the M31 population studies by Henze et al. \citep[see e.g.][]{hen14b}.}. 

The inference then is that LMC 2009a and KT Eri are very similar systems, particularly in terms of their characteristics at eruption.  As well as similarities in SSS development and outburst optical spectra\footnote{\cite{nes13} show that the SSS spectra are however different, with KT Eri and nova V4743 Sgr showing most similarity here, and they discuss inclination as a possible cause of such differences.}, we note that their optical light curves are very similar in form until $\sim100$ days, after which the main difference is the distinct `plateau' observed in KT Eri, lasting until around the time of the re-brightening seen in LMC 2009a see next section. The eruption characteristics of course depend on the WD mass and composition; composition of material undergoing TNR, and mass accretion rate. From Fig.~\ref{fig19} it appears from the observed flux from the disk that $\dot{M}_{\rm acc} $ is also similar in KT Eri and in LMC 2009a. However, the similarity could be caused by the combination of different disk luminosities and inclinations (we note as an aside that for \ KT~Eri, \cite{rib13} find $i = 58^{+6}_{-7}$ deg from  kinematical modelling of the KT~Eri ejecta). It is then interesting to note that the accretion rate in both systems may be similar despite them harboring rather different secondary star types.

\subsection{Optical re-brightening at $\sim250$~days}\label{rebrightening}

\begin{figure}
\includegraphics[width=\columnwidth]{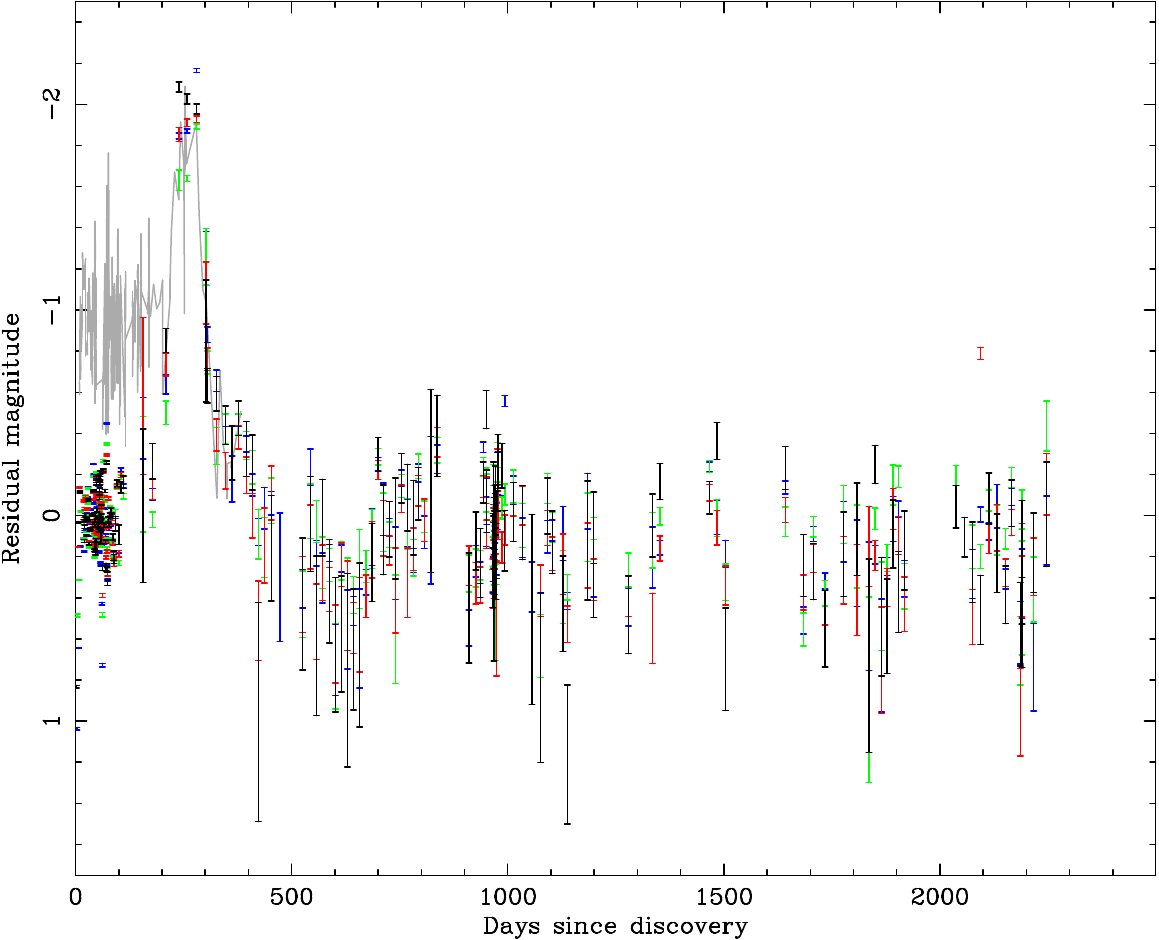}
\caption{SMARTS $B$, $V$, $R$, and $I$ light curves (blue, green, red, and black, respectively) of Nova LMC 2009a following subtraction of the simple light curve model (see Section~\ref{SMARTS_photom}).  The solid gray line is the {\it Swift} UVOT uvw2 lightcurve following similar model subtraction.\label{fig23}}
\end{figure}

Finally, we turn to an unusual feature of the optical light curve. As can be clearly seen in Fig.~\ref{fig2}, in both the $V$- and $I$-band light curves, and the $V$-band residual plot, there is a clear, apparent, re-brightening of the nova around day 250.  In Fig.~\ref{fig23} we show the residuals following a subtraction of a pair of exponential decays and a quiescent level from each of the SMARTS $B$, $V$, $R$, and $I$ light curves (see Section~\ref{SMARTS_photom}).  Here the `bump' is clearly visible, beginning after day $\sim180$ and peaking around day 250.  What is of most interest is perhaps the achromatic nature of this feature, being remarkably similar in morphology and amplitude in all optical bands.  
This feature could of course be easily dismissed as an artefact in the SMARTS data processing due to the low S/N ratio of the data at these late times, however it is also seen in the {\it Swift} uvw2 data, again with a similar morphology and amplitude.

Such features in light curves of explosive transients, with no apparent color dependency (from $\sim0.2$ to $\sim1.25\,\mu$m) are unusual \citep[the secondary maximum seen in the RN T CrB for example appears rather different and was explained, by][as emission from the secondary being heated by a Chandrasekhar mass WD at the Eddington limit and it occurs $\sim1$ orbital period after the eruption]{web87}. If seen unconnected to a transient event, one would be suspicious of a gravitational microlensing event.  A standard \citet{pac86} curve was fitted to the combined $B$, $V$, $R$, and $I$-band data between days 100 and 400 (data outside this range was excluded to avoid any effects due to poor modelling of the nova light-curve behaviour and the 1.2~day periodicity).  This fit to the `bump' was reasonable and returned a time of maximum magnification of day $271\pm1$ (consistent with the sampling of the SMARTS data), and an Einstein radius crossing time (event time-scale) of $108\pm8$ days.  We note that this time-scale is broadly consistent with LMC microlensing events, and that this `bump' would have passed the Paczynski parameter cuts employed by \citet[][see their Table~3, cuts 10 and 11]{wyr11}.
If this `bump' were indeed due to microlensing however, we would expect to see a similar feature in the {\it Swift} X-ray data.  As can be seen in the XRT data (see Fig.~\ref{fig11}) there is some interesting behaviour around day 250, but also there is a gap in the XRT data between day 257 and 279.  The `bump' is also coincident with the start of the X-ray count decreasing and the end of the SSS phase, and the two are most likely connected, but it is hard to reconcile the apparent achromatic nature of the `bump' with the physical processes ongoing in the nova system at that time.

Considering the purely statistical likelihood of this bump being related to a microlensing event, \citet{ben05} estimated the total LMC microlensing rate to be $\Gamma\sim1.5\times10^{-9}$\,events\,source$^{-1}$\,yr$^{-1}$, so clearly the probability of witnessing a microlensing event coincident with a given nova (or any given star for that matter) is negligible, no matter how long one observes that nova.  However, the question we should be asking is, given all the known novae, what is the probability that we witness at least one microlensing event?  Well over 1,500 novae are now known, the majority of these residing in either the Milky Way or M31.  If we assume that both the Galaxy and M31 have the same microlensing rate $\Gamma\sim2\times10^{-5}$\,events\,source$^{-1}$\,yr$^{-1}$ \citep{sum13}, then the probability that one of these systems has been lensed is $\sim3$\,per~cent per year of follow-up. This is low, but not prohibitively so.

\section{Conclusions}

Our optical, UV and X-ray observations of this recurrent nova have led us to the following primary conclusions:

\begin{itemize}

\item
The absolute magnitude of the nova at peak is compatible with that expected from  the Classical Nova MMRD relationship when the relationship was derived from a sample that included such very fast novae.

\item 
There is evidence of a short--lived ($t < 9$ d) high ionisation phase that may in turn be indicative of shocks in the ejecta.

\item
The onset of the SSS phase is preceded by the strengthening of the He~{\sc ii} emission lines, as seen in other novae.

\item
The relatively late emergence and decay of the SSS imply that the WD mass is less than that in either RS Oph or U Sco and, coupled with the observed SSS peak effective temperature, the best estimate is that it lies in the range $1.1 M_\odot \lesssim M_{\rm WD} \lesssim 1.3 M_\odot$.

\item
Periodic oscillations with $P = 1.2$ days are present in both the optical/UV and the X-ray data, with the former leading the latter by $\sim0.3$ days; we interpret this as reflecting the orbital period of the central binary system.

\item
The SSS is initially highly variable, as is the case in several of the other novae observed by {\it Swift}.

\item
We have identified the progenitor system which appears to show emission from a highly luminous accretion disk with a secondary star that is most likely a sub-giant (making LMC 2009a an SG-nova). 

\item
The derived mean accretion rate onto the WD is extremely high ($\dot{\it{M}}_{\rm acc} \simeq 3.6^{+4.7}_{-2.5} \times 10^{-7} M_\odot$ yr$^{-1}$), consistent with observations of the progenitor system, and the recurrent nature of the nova despite the mass of the WD being less than that in e.g.\ RS~Oph or U~Sco.

\item
The slightly lower observed expansion velocities of the ejecta than in some other RNe are consistent with models of the nova explosion employing high accretion rates and WD masses in the range derived above.

\item
LMC 2009a shows several remarkable similarities to the Galactic nova KT Eri and it is suggested that they may be examples of a class of RNe with WDs $\sim 0.1 - 0.3 M_\odot$ away from the Chandrasekhar mass, but with very high inter-eruption mean mass transfer rates. 

\item
There is an intriguing achromatic re-brightening in the optical, near-IR and near-UV light curves at around 250 days which has similarities to a microlensing event, but may be related to the SSS turnoff, and whose true nature is yet to be determined.

\end{itemize}

Over all, LMC 2009a is the best observed confirmed extragalactic RN to-date, with an observational dataset comparable to that of the better observed Galactic examples, and deserves further follow-up. For example, photometry and high resolution spectroscopy at quiescence should be used to investigate further the origin of the 1.2d periodicity and thus to determine more precisely the true nature of the central binary system. As the eruption light curve appears to spend around 1 year above quiescence, and over 100 days above 18th magnitude visually, we encourage optical observers to check their archives for any missed eruptions, particularly between 1971 and 2009. In addition, it would be possible to use the 2009 outburst light curve as a template to identify those epochs where no eruption occurred.

\acknowledgments

This publication makes use of data products from the Two Micron All Sky Survey, which is a joint project of the University of Massachusetts and the Infrared Processing and Analysis Center/California Institute of Technology, funded by the National Aeronautics and Space Administration and the National Science Foundation.  This research has made use of the VizieR catalog access tool, CDS, Strasbourg, France.  APB, JPO, and KLP acknowledge support from the UK Space Agency. SS acknowledges partial support from NSF and NASA grants to Arizona State University. FMW thanks the Provost and Vice President for Research at Stony Brook University for enabling continued participation in the SMARTS partnership; MJD thanks Eamonn Kerrins for useful discussions of the potential microlensing event; we thank the {\it Swift} PI, Neil Gehrels, and operations team for their support in obtaining the {\it Swift} observations, and we are grateful to an anonymous referee whose very insightful comments and suggestions helped to improve the original manuscript.

{\it Facilities:} \facility{SMARTS}, \facility{Swift}.

\end{document}